\documentclass[a4paper]{article}

\usepackage{amssymb}
\usepackage{amsmath}
\usepackage{natbib}
\usepackage{color}
\usepackage{url}
\usepackage{graphicx}

\def\bSig\mathbf{\Sigma}

\newcommand{\tr}{\mbox{tr}}

\def\bfa{\mathbf{a}}
\def\bfb{\mathbf{b}}
\def\bfe{\mathbf{e}}

\def\bfJ{\mathbf{J}}
\def\bfX{\mathbf{X}}
\def\bfV{\mathbf{V}}
\def\bfW{\mathbf{W}}
\def\cL{\mathcal{L}}
\def\tr{^{\mathrm{\scriptsize T}}}
\def\ldotdelta{\dot{\ell}_{\delta}}
\def\ldoteta{\dot{\ell}_{\eta}}
\def\lddotdelta{\ddot{\ell}_{\delta\delta}}
\def\lddotdeltaeta{\ddot{\ell}_{\delta\eta}}
\def\lddoteta{\ddot{\ell}_{\eta\eta}}
\def\bfldotdelta{\boldsymbol{\dot{\ell}}_{\delta}}
\def\bfldoteta{\boldsymbol{\dot{\ell}}_{\eta}}
\def\bflddotdelta{\boldsymbol{\ddot{\ell}}_{\delta\delta}}
\def\bflddotdeltaeta{\boldsymbol{\ddot{\ell}}_{\delta\eta}}
\def\bflddoteta{\boldsymbol{\ddot{\ell}}_{\eta\eta}}
\def\infl{\mathrm{\bf infl}}

\def\ldottheta{\dot{\ell}}
\def\lddottheta{\ddot{\ell}}
\def\bfldottheta{\boldsymbol{\dot{\ell}}}
\def\bflddottheta{\boldsymbol{\ddot{\ell}}}

\newtheorem{theorem}{Theorem}

\title{Modeling continuous monitoring glucose curves \\ by Beta generalized non-parametric models}

\author{Nihan Acar-Denizli
	and 
	Pedro Delicado
	\\
	Department of Statistics and Operations Research, \\
	Universitat Polit\`ecnica de Catalunya - BarcelonaTech (UPC),\\
	Jordi Girona 31, 08034, Barcelona, SPAIN}

\begin{document}
	
\maketitle

\begin{abstract}
We present a functional data analysis approach for studying time-dependent, continuous glucose monitoring data with repeated measures for each individual in an experiment.
After scaling the glucose concentration curves to the interval $[0, 1]$, we model them by using a Beta distribution with two time-varying parameters. 
In this context, we develop a local linear maximum likelihood smoothing procedure that is valid when more than one parameter depends on time.
Our approach requires much fewer observations than previous functional methods for this setting and is also applicable when only one individual (or a few) is available. 
We evaluate the performance of our estimator in terms of computation time and model fit using a synthetic dataset as well as a large, real clinical trial dataset. We also compare our method with existing methods in the literature.
From a methodological point of view, we contribute to extend local likelihood estimation from one to two time-varying parameters by developing theoretical expressions for estimation and for  approximating the leave-one-out cross-validation. 
Moreover, we show that this kernel-based approach competes with spline-based estimation methods, the dominant line of functional regression models today.  
\end{abstract}

\noindent{\bf Keywords:}
Generalized non-parametric models;  Local linear likelihood estimation; Beta functional model; Time-varying parameters; Continuous glucose monitoring.

\section{Introduction}
\label{sec:intro}

Medical devices have an important role in monitoring subject's health related parameters in medicine. Specifically, wearable devices have gained importance in different fields of medicine due to the facility of its use and the allowance of continuous monitoring. 
For instance, the accelerometers are used to monitor movements of subjects during a day, heart rate monitors track electrical activity of heart during sleep or during physical activities and the glucometers are used to measure blood glucose levels (BGL) of patients. 
These devices can take measurements per minute or at 5-minutes intervals. 
Monitoring health related parameters in a continuous time interval facilitates the diagnosis of important diseases. Specifically in continuous glucose monitoring (CGM) observing BGL over time helps to diagnose diabetes \citep[see, e.g.,][]{Gaynanova2022}. 

Since the continuous monitoring brings correlated and big sizes of data, novel statistical methods were proposed to analyze this type of data. 
One of the proposed methods for continuous monitoring data is functional data analysis (FDA) which is based on considering data as a function over a continuous time interval and analyzing random functions rather than random vectors. \citet{Ramsay2005}, \citet{KokRei:2017} and 
\citet{CrGoLeCu:2024}
are excellent introductory books on FDA. They explain the methodological concepts of FDA and summarize different methods to reduce dimension, model and cluster functional data. The number of studies using FDA on wearable device data has increased significantly in recent years, as reviewed in 
\citet{AcarDelicado:2025}. 
See also the example of wearable device data in the book of 
\citet{CrGoLeCu:2024}. 

Among the studies that use FDA to analyze CGM data, \citet{Gaynanova2022} proposed a model based on Beta distribution. The main aim of the study was to analyze within and between subject variability of glucose concentrations of a group of patients with Type 2 diabetes.
To do that, the glucose concentration curves of subjects were first scaled to $[0,1]$ (the range of values of the Beta distribution) by considering subject-specific minimum and maximum BGL values. 
Then, subject-specific mean and variance functions were computed, from which parameter functions of the beta distribution and quantile functions were later derived. 
The present study's main motivation is to propose an alternative estimation method for this time-dependent Beta distribution model.

We consider a simplified version of the model proposed in \citet{Gaynanova2022} assuming independent observations, whose estimation requires a much lower number of observations and can be performed separately for each individual.
We propose fitting a two-parameter generalized non-parametric regression model, which is typically estimated using spline-based techniques (see, for instance, GAMLSS in \citealp{GAMLSS_2024}, and the references therein). In contrast to the splines approach, we address this model estimation using the local likelihood methodology \citep{Loader:1999}. 
Our choice is mainly based on methodological concerns.
Many problems in non-parametric function estimation (from density estimation to non-parametric regression to generalized non-parametric regression) have been approached in parallel using either kernel techniques (including local polynomial fits and local likelihood), or spline basis expansions. 
However, as far as we know, the estimation of a two-parameter generalized non-parametric regression model has only been addressed using B-splines. Thus, we aim to fill this gap. 
We check the practical performance of our proposal, as well as that of competitor approaches, in the clinical trial data set ``REPLACE-BG" \citep{aleppo2017replace}.

The paper is organized as follows. 
In Section \ref{sec:func_Beta}, we give the fundamentals of the Beta functional model, describing first the proposal of \cite{Gaynanova2022} and then introducing our simplified version with independent observations.
In Section \ref{sec:local_likelihood_two}, we explain the extension of maximum local linear likelihood estimation to the case of two parameters. In Section \ref{sec:loc_lik_est_Beta} 
the proposed method is specialized to the Beta distribution and it is
implemented in R. 
In Section \ref{sec:App}, we introduce the real data set on which we apply our approach and interpret the results. 
Finally, Section \ref{sec:concl} summarizes our conclusions.

\section{Beta functional models}\label{sec:func_Beta}
We will divide this section into two parts. In Section \ref{sec:MlBeta_GayPunCra}, we explain the estimation procedure of the multilevel Beta Functional model proposed by \cite{Gaynanova2022}. In Section \ref{sec:MlBeta_simplified}, we propose a simplified model where all observations are independent.

\subsection{Multilevel Beta functional model} 
\label{sec:MlBeta_GayPunCra}
\citet{Gaynanova2022} consider the following repeated measures functional model for CGM data, which they call {\em multilevel Beta functional model}.
Let $G_{ik}(t)$, $t\in[a,b]$, be the functional data recording the BGL at times $t$ in the interval $[a,b]$, corresponding to the $k$-th observation of the $i$-th patient, with $k=1,\ldots,n_i$, and $i=1,\ldots, n$.  
The range of values for $G_{ik}(t)$ is assumed to depend on the individual $i$. Let $[m_i, M_i]$ be this range. 
\citet{Gaynanova2022} assumes that for each individual $i$ the marginal distribution of the BGL data, rescaled to the $[0,1]$ interval,
follows a Beta distribution with parameters depending on $t$ and $i$:
\begin{equation}\label{eq:Multilevel_Beta_Model}
Y_{ik}(t)=\frac{G_{ik}(t)-m_i}{M_i-m_i}
\sim \mbox{Beta}(\alpha_i(t),\beta_i(t)), t\in[a,b],
\end{equation}
for $k=1,\ldots,n_i$ and $i=1,\ldots, n$.
In practice, the functional data $Y_{ik}(t)$ are not observed for all $t\in[a,b]$ but only for times $s_1<\cdots<s_r$ in $[a,b]$ forming a {\em fine grid}: $r$ is {\em large} and 
$\max_{v}(s_{v+1}-s_{v-1})$ 
is {\em small}, where $s_0=a$ and $s_{r+1}=b$.
We assume here that the observation time grid is common for all individuals and all repetitions 
(in \citealp{Gaynanova2022}, more flexibility is allowed when choosing the grid).

In addition to the Beta marginal distribution, in \citet{Gaynanova2022} it is assumed that the functions observed for different individuals are independent and that the different functions observed for the same individual are independent (given $m_i$, $M_i$ and the functional parameters $\alpha_i(t)$ and $\beta_i(t)$). 
However, \citet{Gaynanova2022} do not give any indication about the dependence structure over time for BGL data corresponding to the same observed function at different times.
Indeed, in their concluding section, they mention this point as a topic for further research.

The model estimation procedure proposed by \citet{Gaynanova2022} combines the method of moments and the functional principal component analysis (FPCA; see, e.g., \citealp{Ramsay2005}).
From the properties of the Beta distribution, it follows that
the expected value and variance of $Y_{ik}(t)$ are
\[
\mu_i(t)=\mathbb{E}(Y_{ik}(t))=\frac{\alpha_i(t)}{\alpha_i(t)+\beta_i(t)},\;\;
\sigma^2_i(t)=\mathrm{Var}(Y_{ik}(t))=\frac{\mu_i(t)(1-\mu_i(t))}{\alpha_i(t)+\beta_i(t)+1},
\]
respectively.
Reciprocally,
\begin{equation}\label{eq:a_ms}
\alpha_i(t)=\mu_i(t)\left( \frac{\mu_i(t)\left(1-\mu_i(t)\right)}{\sigma^2_i(t)} - 1\right), 
\end{equation}
and
\begin{equation}\label{eq:b_ms}
\beta_i(t)=\left(1-\mu_i(t)\right)\left( \frac{\mu_i(t)\left(1-\mu_i(t)\right)}{\sigma^2_i(t)} - 1\right).
\end{equation}
The estimation follows these steps:
\begin{enumerate}
\item $m_i$ and $M_i$ are estimated from the subject-specific minimum and maximum BGL values, respectively. 
\item For each individual $i=1,\ldots, n$, consider the pointwise mean function
$\tilde{\mu}_i(t)=(1/n_i)\sum_{k=1}^{n_i} Y_{ik}(t)$, for $t\in [a,b]$.
\item Perform truncated FPCA on the functional data set $\{\tilde{\mu}_i(t): t\in [a,b], i=1,\ldots, n\}$ to get smoother estimations of the mean functions:
\[
\hat{\mu}_ i(t)= \hat{\mu}(t) + \sum_{h=1}^H \hat{\psi}_{ih} \hat{\phi}_h(t),
\]
where $\hat{\mu}(t)$ is the overall mean function, 
$\hat{\phi}_h(t)$ is the $h$-th principal function, 
$\hat{\psi}_{ih}$ is the score of individual $i$ in the $h$-th principal function, and the number of principal components $H$ is chosen to explain a given percentage of variance. 
\item For each individual $i=1,\ldots, n$, consider the pointwise variance function
$\tilde{\sigma}^2_i(t)=
(1/(n_i-1))\sum_{k=1}^{n_i} \left(Y_{ik}(t)-\hat{\mu}_ i(t)\right)^2$, for $t\in [a,b]$.
\item Perform truncated FPCA, as before, on the functional data set 
$\{\tilde{\sigma}^2_i(t): t\in [a,b], i=1,\ldots, n\}$ to get smoother estimations of the variance functions: $\hat{\sigma}^2_i(t)$. 
\item Use $\hat{\mu}_i(t)$, $\hat{\sigma}^2_i(t)$ and equations (\ref{eq:a_ms}) and (\ref{eq:b_ms}) to obtain 
$\hat{\alpha}_i(t)$ and $\hat{\beta}_i(t)$, the 
estimators of $\alpha_i(t)$ and $\beta_i(t)$, respectively.
\end{enumerate}
Among other, \citet{Gaynanova2022} propose to estimate subject-specific pointwise quantiles of glucose functions from $\hat{\alpha}_i(t)$ and $\hat{\beta}_i(t)$.

\subsection{Multilevel Beta functional model with independent observations} 
\label{sec:MlBeta_simplified}
In Section \ref{sec:MlBeta_GayPunCra} we have seen how \citet{Gaynanova2022} propose to estimate the individual functional parameters $\alpha_i(t)$ and $\beta_i(t)$ from observations
\[
Y_{ik}(s_{v}), 
\, v=1,\ldots, r, 
\, k=1,\ldots, n_i, 
\, i=1,\ldots, n,
\]
coming from model (\ref{eq:Multilevel_Beta_Model}), which assumes independence between 
$Y_{ik}(s_v)$ and $Y_{i'k'}(s_{v'})$ when
$i\ne i'$ or when $i=i'$ and $k\ne k'$, but does not specify what happens for $i=i'$, $k=k'$ and $v\ne v'$. 

Here, we propose a simplified version of the multilevel Beta functional model (\ref{eq:Multilevel_Beta_Model}), assuming independence among all observations.
This model has two main advantages over the one in \citet{Gaynanova2022}. First, the functional parameters $\alpha_i(t)$ and $\beta_i(t)$ can be estimated separately for each individual. Second, much less data is required for the estimation.

To be specific, the {\em multilevel Beta functional model with independent observations} states that, for each individual $i$, 
\begin{equation}\label{eq:MlBeta_simplified}
Y_i(t_{ij}) \sim \mbox{Beta}(\alpha_i(t_{ij}), \beta_i(t_{ij})), \, j=1,\ldots, m_i,
\end{equation}
are independently observed. Independence between different individuals is also assumed. 
The time values $t_{ij}$ are in the fine grid $\{s_1,\ldots, s_r\}$, and can be randomly chosen ({\em random design}) or they can be fixed values known in advance ({\em fixed design}). 

Observe that model (\ref{eq:MlBeta_simplified}), when stated separately for individual $i$, constitutes a {\em generalized non-parametric regression model}, that is, a generalized additive model (GAM) with only one explanatory variable $T$ and response $Y$, where the conditional distribution $(Y|T=t)$ follows a known parametric model with two parameters that vary with $t$ in a non-parametric way. 

The standard approach to GAMs (see, for instance, \citealt{hastie1990generalized},   \citealt{ruppert2003semiparametric} or \citealt{Wood:2017}) only considers one parameter that depends on the explanatory variables.
Particularly relevant is the R implementation of GAM developed by Simon Wood in the library {\tt mgcv}.

An alternative approach is GAMLSS (Generalized Additive Models for Location, Scale and Shape), which allows several parameters to depend on explanatory variables simultaneously. 
\cite{GAMLSS_2005_JRSSC} was the seminal paper on GAMLSS.
Several books (\citealp{GAMLSS_2017, GAMLSS_2019,GAMLSS_2024}) have been written later by the same research team, covering relevant extensions of GAMLSS, their implementation in R (library {\tt gamlss}), and several real data applications. See also the GAMLSS project website \url{https://www.gamlss.com/}.

The implementation of both GAM and GAMLSS typically involves expanding the parameter functions in a basis of splines. This way, the non-parametric model is managed as a parametric one (where the parameters are the coefficients in the spline expansions). Estimation is made by penalized maximum likelihood, where the penalty term prevents overfitting. 
This approach requires setting a {\em tuning parameter}, which controls the relative weight of the penalty term in the maximized objective function. 

In non-parametric regression, local polynomial fitting is an alternative classical approach to basis expansion. 
The idea is to fit a simple model (a low-degree polynomial model, usually a linear model or even a constant) locally around the point $t_0$ at which we want to estimate the regression function $\mathbb{E}(Y|T=t_0)$. Localization around $t_0$ is achieved by assigning weights to the observed data $(t_j,Y_j)$ by a {\em kernel function} $K((t_j-t_0)/h)$, where $K$ is a symmetric function, typically non-negative, with a maximum at $0$. 
The tuning parameter $h$ (known as {\em bandwidth}) controls the flexibility of the model. Too much flexibility implies overfitting, while too little leads to a poor fit.

Classical references on local polynomials (also referred to as {\em kernel methods}) are \cite{WandJones:1995} or  \cite{FanGijbels:1996}. See also Chapters 5 and 6 in \cite{HastieTibshiraniFriedman:2009}, devoted to spline smoothing and kernel methods, respectively, where a comparison of both methods can be found. 
It is worth mentioning the book of \cite{Loader:1999}, in which the kernel approach to generalized nonparametric models is presented in detail under the name {\em local likelihood estimation}.
Note that \cite{Loader:1999} only considered the case of one parameter depending on the explanatory variable, 
while two years before, in \cite{Aerts_Claeskens:1997}, the local polynomial fitting was extended to multiparameter likelihood models. In particular, \cite{Aerts_Claeskens:1997} proved the consistency and asymptotic normality of local polynomial likelihood estimators under typical likelihood regularity conditions and smoothness conditions on the parameters as functions of $t$.

In this context, we propose a local likelihood approach to estimate the multilevel Beta functional model with independent observations.
In Section \ref{sec:local_likelihood_two} we extend some aspects of the local likelihood methodology (given in \citealp{Loader:1999}, for the one-parameter case and not covered in \citealp{Aerts_Claeskens:1997}, for the multiparameter setup) to the case of two-parameter models, such as the beta distribution.
To be specific, in one-parameter local likelihood estimation \cite{Loader:1999} 
gives the expressions for the {\em influence function}, for the {\em effective degrees of freedom} and for a {\em generalization of the Akaike information creiterion}. 
These elements enable the automatic selection of the bandwidth $h$ based on an approximation of the leave-one-out cross-validation method. 
This is expected to result in considerable savings in computing time.
For the sake of completeness, the Appendix \ref{app:loader} summarizes Chapter 4 in \cite{Loader:1999}, in which these components were formulated.

Once the multilevel Beta functional model (\ref{eq:MlBeta_simplified}) is estimated (by either GAM, GAMLSS or local likelihood approaches),
a Beta distribution with time-varying parameters 
$\{\widehat{\mbox{Beta}}_i(t)\equiv \mbox{Beta}(\hat{\alpha}_i(t),\hat{\beta}_i(t)):t\in[a,b]\}$
is obtained for each individual $i=1,\ldots,n$, in the sample.
Then a further step would be required to summarize the information contained in these individual estimations. 
For this reason, we propose to apply dimensionality reduction techniques 
to have a global view of the complete dataset. 
See Section \ref{sec:App_MDS_FPCA} for a real data example.

\section{Local likelihood in a two-parameters model} \label{sec:local_likelihood_two}
Without loss of generality, from now on we consider the time interval to be $[0,1]$. Furthermore, since the estimation is done separately for each individual, we ignore the individual subscript $i$.
Let $Y_1,\ldots, Y_m$ be independent random variables with probability
density (or mass) function $f(y;\delta(t_j),\eta(t_j))$, $j=1,\ldots, m$, where $0<t_1<\cdots<t_m < 1$ are known constants, 
$\delta(t)$ and $\eta(t)$ are {\em smooth} functions of $t$, and they have no restrictions ($\delta(t)$ and $\eta(t)$ can take any value in $\mathbb{R}$).  
In the Beta functional model, we take
$\delta(t)=\log(\alpha(t))$ and $\eta(t)=\log(\beta(t))$.
Let $y_1,\ldots, y_m$ be the observed values of $Y_1,\ldots, Y_m$.

The objective of local likelihood estimation is to estimate the functions $(\delta(t),\eta(t))$, for all $t\in[0,1]$, in a non-parametric way using the kernel approach.  
We follow \cite{Loader:1999}, where the one-dimensional parameter case is developed in detail. 
We focus on estimating $(\delta(t),\eta(t))$ by maximum local linear likelihood. 
A related approach can be found in \cite{AcarDelicado:2025:IWFOS}, where local constant maximum likelihood is used instead. 

\subsection{Maximum local linear log-likelihood estimation}
Let $t_0$ be a point in $[0,1]$ at which we want to estimate the values $(\delta(t_0),\eta(t_0))$. 
For $t$ in a neighborhood of $t_0$, consider the first-order Taylor approximations
\[
\begin{array}{lcl}
	\delta(t)&\approx& 
	\delta_{t_0}(t)=\delta(t_0) + \delta'(t_0)(t-t_0)\equiv a_0 + a_1 (t-t_0)
	=\langle \bfa, A(t-t_0) \rangle, \\
	\eta(t)&\approx& 
	\eta_{t_0}(t)=\eta(t_0) + \eta'(t_0)(t-t_0) \equiv b_0 + b_1 (t-t_0)
	=\langle \bfb, A(t-t_0) \rangle, 
\end{array}
\]
where $\bfa=(a_0, a_1)\tr$, $\bfb=(b_0,b_1)\tr$ and $A(u)=(1,u)\tr$, for $u\in \mathbb{R}$.
The local linear log-likelihood function around $t_0$ is defined as
\begin{equation}\label{eq:llll_function}
	\cL_{t_0}(\bfa,\bfb)=\sum_{j=1}^m w_j(t_0) 
	\ell(y_j;\langle \bfa, A(t_j-t_0) \rangle, \langle \bfb, A(t_j-t_0) \rangle),
\end{equation}
where $\ell(y;\delta,\eta)=\log\left(f(y;\delta,\eta)\right)$, and the weights $w_j(t_0)$ are given by a kernel function $K$ and a bandwidth parameter $h$: $w_j(t_0)=K((t_j-t_0)/h)$. 

Let $(\hat{\bfa},\hat{\bfb})$ be the maximum local linear likelihood estimators of $(\bfa,\bfb)$:
\begin{equation}\label{eq:max_llll_function}
(\hat{\bfa},\hat{\bfb}) =\arg\max_{(\bfa,\bfb)} \cL_{t_0}(\bfa,\bfb).
\end{equation}
Observe that $(\hat{\bfa},\hat{\bfb})$ depends on $t_0$. 
The local linear likelihood estimators of $\delta(t_0)$ and $\eta(t_0)$ are defined as
\[
\hat{\delta}(t_0)= 
\langle \hat{\bfa}, A(0) \rangle=\hat{a}_0, \,\,
\hat{\eta}(t_0)= 
\langle \hat{\bfb}, A(0) \rangle=\hat{b}_0.
\]

In Section \ref{sec:first_order_condts}, we provide an operative expression for the gradient of $\cL_{t_0}(\bfa,\bfb)$, which is useful for solving the problem (\ref{eq:max_llll_function}) numerically and for approximating the leave-one-out bandwidth choice methods discussed in Section \ref{sec:BandwidthChoice}. 

%

\subsection{First-order conditions}\label{sec:first_order_condts}
The first-order conditions (also known as the estimating equations) for the maximum local linear likelihood problem (\ref{eq:max_llll_function}) around a generic $t$ are 
\[
\nabla_{(\bfa,\bfb)} \cL_{t}(\bfa,\bfb)=\mathbf{0}_4,
\]
where $\mathbf{0}_d$ is the vector of zeros in $\mathbb{R}^d$. 

Let $\nabla_{\bfa} \cL_{t}(\bfa,\bfb)$ (respectively, $\nabla_{\bfb} \cL_{t}(\bfa,\bfb)$) be the first (resp., last) two components of $\nabla_{(\bfa,\bfb)} \cL_{t}(\bfa,\bfb)$. 
Then, taking into account the definition of $\cL_t$ in equation (\ref{eq:llll_function}),
\[
\nabla_{\bfa} \cL_{t}(\bfa,\bfb) = 
\sum_{j=1}^m w_j(t) A(t_j-t) 
\ldotdelta(y_j;\langle \bfa, A(t_j-t) \rangle, \langle \bfb, A(t_j-t) \rangle)
=
\]
\[
\bfX\tr \bfW \bfldotdelta(\bfX \bfa, \bfX \bfb), 
\] 
where 
$\bfX$ is the $m\times 2$ matrix with $j$-th row $A(t_j-t)\tr=(1,t_j-t)$, 
$\bfW$ is the $m\times m$ diagonal matrix with $j$-th element in the diagonal $w_j(t)$, and 
$\bfldotdelta(\bfX \bfa, \bfX \bfb)$ is the column vector in $\mathbb{R}^m$  with $j$-th element 
$\ldotdelta(y_j;\langle \bfa, A(t_j-t) \rangle, \langle \bfb, A(t_j-t) \rangle)$. 
Analogously,
\[
\nabla_{\bfb} \cL_{t}(\bfa,\bfb) = 
\sum_{j=1}^m w_j(t) A(t_j-t) 
\ldoteta(y_j;\langle \bfa, A(t_j-t) \rangle, \langle \bfb, A(t_j-t) \rangle)
=
\]
\[
\bfX\tr \bfW \bfldoteta(\bfX \bfa, \bfX \bfb). 
\]
Therefore, the estimating equations can be written in matrix notation as
\[
\left.
\begin{array}{c}
	\bfX\tr \bfW \bfldotdelta(\bfX \bfa, \bfX \bfb)= 
	\mathbf{0}_2
	\\
	\bfX\tr \bfW \bfldoteta(\bfX \bfa, \bfX \bfb)= 
	\mathbf{0}_2
\end{array}
\right\}
\]

\subsection{Bandwidth choice}\label{sec:BandwidthChoice}
We have extended to two-parameter distributions the approximation method for leave-one-out developed in \cite{Loader:1999} for the one-parameter case (see Appendix \ref{app:loader}).

\subsubsection{Leave-one-out cross-validation}\label{sec:looCV}
For a particular value of the bandwidth $h$, the leave-one-out cross-validation version of the log-likelihood function is 
\begin{equation}\label{eq:loo_log_lik}
	\text{CV}(h) = \sum_{j=1}^{m}
	\ell(y_j; \hat{\delta}_{(j)}(t_j),\hat{\eta}_{(j)}(t_j)),
\end{equation}
where $(\hat{\delta}_{(j)}(t), \hat{\eta}_{(j)}(t))$ are the estimated parameter functions when the $j$-th observation has been left out of the sample and a local linear maximum likelihood estimator has been used with bandwidth $h$. 
Then, $h$ is chosen to maximize $\text{CV}(h)$:
\[
h_{\text{\scriptsize CV}} = \arg \max_h \text{CV}(h).
\]

This bandwidth selector requires to estimate $m$ parameter functions (where $m$ is the number of observed pairs $(t_j, Y_j)$), each of them involving $(m-1)$ maximum local likelihood estimations. This implies a high computational cost.
In certain cases (linear estimators of the regression functions and, in particular, linear smoothers) it is possible to deduce an exact expression for the $j$-th term in (\ref{eq:loo_log_lik}) from the corresponding one in the full log-likelihood function  
\[
\sum_{j=1}^{m}
\ell(y_j;\hat{\delta}(t_j), \hat{\eta}(t_j)),
\]
where $(\hat{\delta}(t), \hat{\eta}(t))$ are estimated with the complete data set. 
In this case, the parameter functions have to be estimated only once. 
When an exact expression is unavailable, approximations can speed up the computation of (\ref{eq:loo_log_lik}). 

\subsubsection{Approximate leave-one-out cross-validation}\label{sec:Approx_looCV}
We develop now an approximate expression for (\ref{eq:loo_log_lik}) in the case of a two-parameter model local linear likelihood estimation.
We follow the same steps as in \cite{Loader:1999}, summarized in the Appendix \ref{app:loader}.

For $t=t_j$ and for $\lambda\in [0,1]$, consider the modified estimating equations 
\[
\left.
\begin{array}{c}
	\bfX\tr \bfW \bfldotdelta(\bfX \bfa, \bfX \bfb)
	- \lambda K(0) A(0)  
	\ldotdelta(y_j;\langle \bfa, A(0) \rangle, \langle \bfb, A(0) \rangle) 
	= 
	\mathbf{0}_2
	\\
	\bfX\tr \bfW \bfldoteta(\bfX \bfa, \bfX \bfb)
	- \lambda K(0) A(0) 
	\ldoteta(y_j;\langle \bfa, A(0) \rangle, \langle \bfb, A(0) \rangle) 
	= 
	\mathbf{0}_2
\end{array}
\right\}
\]
and let $(\hat{\bfa}(\lambda),\hat{\bfb}(\lambda))$ be the solution. 
It follows that $(\hat{\bfa}(0),\hat{\bfb}(0))$ is the maximum local likelihood estimate of 
$(\bfa,\bfb)$ at $t=t_j$, and that 
$(\hat{\bfa}(1),\hat{\bfb}(1))$ is the leave-one-out counterpart. 
We consider the first-order Taylor approximation of $(\hat{\bfa}(1),\hat{\bfb}(1))$ around
$(\hat{\bfa}(0),\hat{\bfb}(0))$:
\[
(\hat{\bfa}(1)\tr,\hat{\bfb}(1)\tr)\tr
\approx
(\hat{\bfa}(0)\tr,\hat{\bfb}(0)\tr)\tr
+ 
\left. 
\frac{d \, (\hat{\bfa}(\lambda)\tr,\hat{\bfb}(\lambda)\tr)\tr}{d\, \lambda}
\right|_{\lambda=0}.
\]
To compute the $4$-dimensional directional derivative with respect to $\lambda$ we write the modified estimating equations evaluated at the solution $(\hat{\bfa}(\lambda),\hat{\bfb}(\lambda))$,
\[
\left(
\begin{array}{c}
	\bfX\tr \bfW \bfldotdelta(\bfX \hat{\bfa}(\lambda), \bfX \hat{\bfb}(\lambda)) \\
	\bfX\tr \bfW \bfldoteta(\bfX \hat{\bfa}(\lambda), \bfX \hat{\bfb}(\lambda))	
\end{array} 
\right)
=
\lambda K(0)
\left(
\begin{array}{c}
	A(0)  
	\ldotdelta(y_j;\langle \hat{\bfa}(\lambda), A(0) \rangle, \langle \hat{\bfb}(\lambda), A(0) \rangle)
	\\
	A(0)  
	\ldoteta(y_j;\langle \hat{\bfa}(\lambda), A(0) \rangle, \langle \hat{\bfb}(\lambda), A(0) \rangle)
\end{array}
\right),
\]
and take derivatives with respect to $\lambda$ at both sides:
\[
\left(
\begin{array}{c}
	\bfX\tr \bfW 
	\left[ 
	\bflddotdelta(\bfX \hat{\bfa}(\lambda), \bfX \hat{\bfb}(\lambda)) \bfX 
	\frac{d \, \hat{\bfa}(\lambda)}{d\, \lambda}
	+
	\bflddotdeltaeta(\bfX \hat{\bfa}(\lambda), \bfX \hat{\bfb}(\lambda)) \bfX
	\frac{d \, \hat{\bfb}(\lambda)}{d\, \lambda}
	\right] 
	\\[.2cm]
	\bfX\tr \bfW 
	\left[ 
	\bflddotdeltaeta(\bfX \hat{\bfa}(\lambda), \bfX \hat{\bfb}(\lambda)) \bfX 
	\frac{d \, \hat{\bfa}(\lambda)}{d\, \lambda}
	+
	\bflddoteta(\bfX \hat{\bfa}(\lambda), \bfX \hat{\bfb}(\lambda)) \bfX
	\frac{d \, \hat{\bfb}(\lambda)}{d\, \lambda}
	\right] 
\end{array} 
\right)
=
\]
\[
K(0)
\left(
\begin{array}{c}
	A(0)  
	\ldotdelta(y_j;\langle \hat{\bfa}(\lambda), A(0) \rangle, \langle \hat{\bfb}(\lambda), A(0) \rangle)
	\\
	A(0)  
	\ldoteta(y_j;\langle \hat{\bfa}(\lambda), A(0) \rangle, \langle \hat{\bfb}(\lambda), A(0) \rangle)
\end{array}
\right)
+
\lambda K(0)
\]
\[
{\scriptsize
	\left(
	\begin{array}{c}
		A(0) \left[  
		\lddotdelta(y_j;\langle \hat{\bfa}(\lambda), A(0) \rangle, 
		\langle \hat{\bfb}(\lambda), A(0) \rangle)  
		\frac{d \, \hat{\bfa}(\lambda)\tr}{d\, \lambda}
		+    
		\lddotdeltaeta(y_j;\langle \hat{\bfa}(\lambda), A(0) \rangle, 
		\langle \hat{\bfb}(\lambda), A(0) \rangle) 
		\frac{d \, \hat{\bfb}(\lambda)\tr}{d\, \lambda}
		\right] A(0) 
		\\[.2cm]
		A(0) \left[  
		\lddotdeltaeta(y_j;\langle \hat{\bfa}(\lambda), A(0) \rangle, 
		\langle \hat{\bfb}(\lambda), A(0) \rangle)  
		\frac{d \, \hat{\bfa}(\lambda)\tr}{d\, \lambda}
		+    
		\lddoteta(y_j;\langle \hat{\bfa}(\lambda), A(0) \rangle, 
		\langle \hat{\bfb}(\lambda), A(0) \rangle) 
		\frac{d \, \hat{\bfb}(\lambda)\tr}{d\, \lambda}
		\right] A(0) 
	\end{array}
	\right),
}
\]
where the terms $\bflddotdelta$, $\bflddotdeltaeta$ and $\bflddoteta$ are $m\times m$ diagonal matrices with second derivatives computed from the elements in $\bfldotdelta$ and $\bfldoteta$.
Now we evaluate the previous equation at $\lambda=0$ and solve in the 4-dimensional directional derivative. 
To simplify the notation, let us define the $(2m)\times (2m)$ matrix
\[
\bfV=
- \left(
\begin{array}{cc}
	\bflddotdelta(\bfX \hat{\bfa}(0), \bfX \hat{\bfb}(0)) &
	\bflddotdeltaeta(\bfX \hat{\bfa}(0), \bfX \hat{\bfb}(0))
	\\
	\bflddotdeltaeta(\bfX \hat{\bfa}(0), \bfX \hat{\bfb}(0)) &
	\bflddoteta(\bfX \hat{\bfa}(0), \bfX \hat{\bfb}(0))
\end{array}\right)
\]
and the $4\times 4$ matrix 
\begin{equation}\label{eq:J_matrix}    
\bfJ= 
\left(
\begin{array}{cc}
	\bfX\tr & \mathbf{0}_{2\times m} \\
	\mathbf{0}_{2\times m} & \bfX\tr 
\end{array}
\right) 
\left(
\begin{array}{cc}
	\bfW & \mathbf{0}_{m\times m} \\
	\mathbf{0}_{m\times m} & \bfW 
\end{array}
\right) 
\bfV 
\left(
\begin{array}{cc}
	\bfX & \mathbf{0}_{m\times 2} \\
	\mathbf{0}_{m\times 2} & \bfX 
\end{array}
\right).
\end{equation}
Under the standard assumption that $\bfW\bfX$ has full rank, the concavity of the likelihood function implies that $\bfJ$ is positive definite 
\cite[see][Theorem 4.1]{Loader:1999}, so it is invertible. 
Observe that $\bfJ$ is obtained by taking the partial derivatives of minus the gradient $\nabla_{(\bfa,\bfb)} \cL_{t}(\bfa,\bfb)$ computed in Section \ref{sec:first_order_condts}, so $-\bfJ$ is the Hessian matrix of the objective function $\cL_{t}(\bfa,\bfb)$ in problem (\ref{eq:max_llll_function}), evaluated at the local maximum likelihood estimator of $(\delta(t),\eta(t))$. 
The Hessian matrix at a generic point $(\delta,\eta)$ can be calculated using similar arguments.
Define also $\bfe_i$, $i=1,\ldots,4$, as the vector of $\mathbb{R}^4$ with a one in the $i$-th position and zeros elsewhere.
Then, the derivatives with respect to $\lambda$ of the modified estimating equations, evaluated at $\lambda=0$, can be expressed as
\[
- \bfJ 
\left. 
\frac{d \, (\hat{\bfa}(\lambda)\tr,\hat{\bfb}(\lambda)\tr)\tr}{d\, \lambda}
\right|_{\lambda=0}
=
K(0) 
(\bfe_1,\bfe_3)
\left(
\begin{array}{c}
	\ldotdelta(y_j;\langle \hat{\bfa}(0), A(0) \rangle, \langle \hat{\bfb}(0), A(0) \rangle)
	\\
	\ldoteta(y_j;\langle \hat{\bfa}(0), A(0) \rangle, \langle \hat{\bfb}(0), A(0) \rangle)
\end{array}
\right),
\]
and it follows that
\[
\left. 
\frac{d \, (\hat{\bfa}(\lambda)\tr,\hat{\bfb}(\lambda)\tr)\tr}{d\, \lambda}
\right|_{\lambda=0}
=
- K(0) 
\bfJ^{-1} 
(\bfe_1,\bfe_3)
\left(
\begin{array}{c}
	\ldotdelta(y_j;\langle \hat{\bfa}(0), A(0) \rangle, \langle \hat{\bfb}(0), A(0) \rangle)
	\\
	\ldoteta(y_j;\langle \hat{\bfa}0), A(0) \rangle, \langle \hat{\bfb}(0), A(0) \rangle)
\end{array}
\right).
\]
Therefore,
\[
{\hat{\delta}_{(j)}(t_j) \choose \hat{\eta}_{(j)}(t_j)}
=
{\bfe_1\tr \choose \bfe_3\tr}
{\hat{\bfa}(1) \choose \hat{\bfb}(1)}
\approx 
\]
\[
{\bfe_1\tr \choose \bfe_3\tr}
{\hat{\bfa}(0) \choose \hat{\bfb}(0)}
-
K(0) 
{\bfe_1\tr \choose \bfe_3\tr}
\bfJ^{-1} 
(\bfe_1,\bfe_3)
\left(
\begin{array}{c}
	\ldotdelta(y_j;\langle \hat{\bfa}(0), A(0) \rangle, \langle \hat{\bfb}(0), A(0) \rangle)
	\\
	\ldoteta(y_j;\langle \hat{\bfa}(0), A(0) \rangle, \langle \hat{\bfb}(0), A(0) \rangle)
\end{array}
\right)
=
\]
\[
{\hat{\delta}(t_j) \choose \hat{\eta}(t_j)}
-
K(0) 
{\bfe_1\tr \choose \bfe_3\tr}
\bfJ^{-1} 
(\bfe_1,\bfe_3)
\left(
\begin{array}{c}
	\ldotdelta(y_j;\hat{\delta}(t_j),\hat{\eta}(t_j))
	\\
	\ldoteta(y_j;\hat{\delta}(t_j),\hat{\eta}(t_j))
\end{array}
\right).
\]
Following the notation used in \cite{Loader:1999},
we define the influence function at $t_j$ as the $2\times 2$ matrix
\[
\infl(t_j) = K(0) 
{\bfe_1\tr \choose \bfe_3\tr}
\bfJ^{-1} 
(\bfe_1,\bfe_3).
\]
Then,
\[
{\hat{\delta}_{(j)}(t_j) \choose \hat{\eta}_{(j)}(t_j)}
\approx 
{\hat{\delta}(t_j) \choose \hat{\eta}(t_j)}
-
\infl(t_j)
\left(
\begin{array}{c}
	\ldotdelta(y_j;\hat{\delta}(t_j),\hat{\eta}(t_j))
	\\
	\ldoteta(y_j;\hat{\delta}(t_j),\hat{\eta}(t_j))
\end{array}
\right).
\] 
Now we approximate $\ell(y_j;\hat{\delta}_{(j)}(t_j), \hat{\eta}_{(j)}(t_j))$ with a first-order Taylor expansion:
\[
\ell(y_j;\hat{\delta}_{(j)}(t_j), \hat{\eta}_{(j)}(t_j))
\approx 
\ell(y_j;\hat{\delta}(t_j), \hat{\eta}(t_j))
-
\]
\[
(\ldotdelta(y_j;\hat{\delta}(t_j),\hat{\eta}(t_j)),
\ldoteta(y_j;\hat{\delta}(t_j),\hat{\eta}(t_j)))
\, 
\infl(t_j)
\left(
\begin{array}{c}
	\ldotdelta(y_j;\hat{\delta}(t_j),\hat{\eta}(t_j))
	\\
	\ldoteta(y_j;\hat{\delta}(t_j),\hat{\eta}(t_j))
\end{array}
\right).
\] 
This is precisely the type of approach we anticipated at the end of Section \ref{sec:looCV}.
We have proven the following theorem.

\begin{theorem}\label{th:CV}
	\[
	\text{CV}(h) = 
	\sum_{j=1}^{m}
	\ell(y_j;\hat{\delta}_{(j)}(t_j), \hat{\eta}_{(j)}(t_j))
	\approx 
	\widetilde{\text{CV}}(h) = 
	\sum_{j=1}^{m} 
	\ell(y_j;\hat{\delta}(t_j), \hat{\eta}(t_j))
	- 
	\]
	\[
	\sum_{j=1}^{m}
	(\ldotdelta(y_j;\hat{\delta}(t_j),\hat{\eta}(t_j)),
	\ldoteta(y_j;\hat{\delta}(t_j),\hat{\eta}(t_j)))
	\, 
	\infl(t_j)
	\left(
	\begin{array}{c}
		\ldotdelta(y_j;\hat{\delta}(t_j),\hat{\eta}(t_j))
		\\
		\ldoteta(y_j;\hat{\delta}(t_j),\hat{\eta}(t_j))
	\end{array}
	\right).
	\]
\end{theorem}

As a consequence of the previous result, the bandwidth $h$ can be chosen to maximize $\widetilde{\text{CV}}(h)$:
\[
h_{\widetilde{\text{\scriptsize CV}}} = \arg \max_h \widetilde{\text{CV}}(h).
\]

Following \cite{Loader:1999}, a definition of the {\em effective degrees of freedom} of the estimation can be derived from Theorem \ref{th:CV}. 
Remember that the deviance is defined as minus twice the log likelihood of the fitted model. Then, the deviance computed from the leave-one-out fitted model is
\[
\mathrm{Dev}_{\mathrm loo}(h)
\approx  
\mathrm{Dev}(h) 
+  
\]
\[
2 \sum_{j=1}^{m}
(\ldotdelta(y_j;\hat{\delta}(t_j),\hat{\eta}(t_j)),
\ldoteta(y_j;\hat{\delta}(t_j),\hat{\eta}(t_j)))
\, 
\infl(t_j)
\left(
\begin{array}{c}
	\ldotdelta(y_j;\hat{\delta}(t_j),\hat{\eta}(t_j))
	\\
	\ldoteta(y_j;\hat{\delta}(t_j),\hat{\eta}(t_j))
\end{array}
\right).
\]
Defining the effective degrees of freedom as
\[
\nu =
\sum_{j=1}^{m}
(\ldotdelta(y_j;\hat{\delta}(t_j),\hat{\eta}(t_j)),
\ldoteta(y_j;\hat{\delta}(t_j),\hat{\eta}(t_j)))
\, 
\infl(t_j)
\left(
\begin{array}{c}
	\ldotdelta(y_j;\hat{\delta}(t_j),\hat{\eta}(t_j))
	\\
	\ldoteta(y_j;\hat{\delta}(t_j),\hat{\eta}(t_j))
\end{array}
\right)
\]
we obtain a generalization of the Akaike information criterion:
\[
\mathrm{AIC}(h)=\mathrm{Dev}(h) 
+ 2 \nu.
\]
Therefore, 
\[
h_{\widetilde{\text{\scriptsize CV}}} = 
\arg \max_h \widetilde{\text{CV}}(h) =
\arg \min_h \mathrm{AIC}(h).
\]
The main argument of \cite{Loader:1999} for computing $h_{\widetilde{\text{\scriptsize CV}}}$ as an approximation of 
$h_{\text{\scriptsize CV}}$ is that the computing time required for $h_{\widetilde{\text{\scriptsize CV}}}$ is much lower than that for 
$h_{\text{\scriptsize CV}}$. 
We expect the two-parameter setting to have a similar advantage in computation time. In Section \ref{sec:toy_example}, whether or not this is the case will be checked.

\section{Local likelihood estimation in the Beta model} \label{sec:loc_lik_est_Beta}
The previous concepts and results are now applied to the Beta distribution, having density function
\[
\xi(y;\alpha,\beta)=\frac{\Gamma(\alpha+\beta)}{\Gamma(\alpha)\Gamma(\alpha)}y^{\alpha}(1-y)^{\beta}, \, y \in [0,1].
\]
Let $\rho(y;\alpha,\beta)$ be its logarithm:
\[
\rho(y;\alpha,\beta)=
\log \Gamma(\alpha+\beta)
-\log \Gamma(\alpha) - \log\Gamma(\alpha)
+ \alpha \log y + \beta \log (1-y), \, y \in [0,1].
\]
Then, $f(y;\delta,\eta)=\xi(y;e^\delta,e^\eta)$
and 
$\ell(y;\delta,\eta)=\rho(y;e^\delta,e^\eta)$.
Let us compute the first and second partial derivatives of $\ell(y;\delta,\eta)$ with respect to $(\delta,\eta)$:
\[
\ldotdelta(y;\delta,\eta)\equiv
\frac{\partial \rho(y;e^\delta,e^\eta)}{\partial \delta}=
\left( \psi(e^{\delta}+e^{\eta}) - \psi(e^{\delta}) + \log y\right) e^{\delta},
\]
\[
\ldoteta(y;\delta,\eta)\equiv 
\frac{\partial \rho(y;e^\delta,e^\eta)}{\partial \eta}=
\left( \psi(e^{\delta}+e^{\eta}) - \psi(e^{\eta}) + \log (1-y)\right) e^{\eta},
\]
\[
\lddotdelta(y;\delta,\eta)\equiv
\frac{\partial^2 \rho(y;e^\delta,e^\eta)}{\partial \delta^2}=
\]
\[
\left( \psi^{(1)}(e^{\delta}+e^{\eta}) - \psi^{(1)}(e^{\delta})\right) e^{2\delta} +
\left( \psi(e^{\delta}+e^{\eta}) - \psi(e^{\delta}) + \log y\right) e^{\delta},
\]
\[
\lddoteta(y;\delta,\eta)\equiv
\frac{\partial^2 \rho(y;e^\delta,e^\eta)}{\partial \eta^2}=
\]
\[
\left( \psi^{(1)}(e^{\delta}+e^{\eta}) - \psi^{(1)}(e^{\eta})\right) e^{2\eta} +
\left( \psi(e^{\delta}+e^{\eta}) - \psi(e^{\eta}) + \log y\right) e^{\eta},
\]
\[
\lddotdeltaeta(y;\delta,\eta)\equiv
\frac{\partial^2 \rho(y;e^\delta,e^\eta)}{\partial \delta\partial \eta}=
\psi^{(1)}(e^{\delta}+e^{\eta})  e^{\delta}e^{\eta},
\]
where $\psi$ is the digamma function (the derivative of the logarithm of the gamma function) and $\psi^{(1)}$ is the trigamma function (the derivative of the digamma).



\subsection{Implementation} \label{sec:Implementation}
In accordance with the preceding expressions, the Beta model local likelihood estimation has been implemented in R, as outlined in Section \ref{sec:local_likelihood_two}.
To be specific, the optimization problem (\ref{eq:max_llll_function}) is solved by using either the non-linear minimization function {\tt nlm} (which uses a Newton-type algorithm) or the general-purpose optimization function {\tt optim} (using either the {\tt Nelder-Mead} downhill simplex method or the {\tt BFGS} quasi-Newton method). 
When using {\tt nlm}, we provide it with the gradient $\nabla_{(\bfa,\bfb)} \cL_{t}(\bfa,\bfb)$ and Hessian 
$-\bfJ$ of $\cL_{t}(\bfa,\bfb)$, which are computed as indicated in Sections \ref{sec:first_order_condts} and \ref{sec:Approx_looCV}, respectively.
If {\tt optim} is used instead, the gradient $\nabla_{(\bfa,\bfb)} \cL_{t}(\bfa,\bfb)$ is provided when the specified optimization method is {\tt BFGS}.

Regarding the bandwidth choice, three different methods have been implemented:
leave-one-out cross-validation (Section \ref{sec:looCV}), 
approximate leave-one-out cross-validation (Section \ref{sec:Approx_looCV}),
and $k$-fold cross-validation, which maximizes
\begin{equation}\label{eq:kfoldCV}
	\text{CV}(h) = \sum_{a=1}^{k} \sum_{j\in A_a}
	\ell(y_j; \hat{\delta}_{(A_a)}(t_j),\hat{\eta}_{(A_a)}(t_j)),
\end{equation}
where $A_1,\ldots,A_k$ is a random partition of $\{1,\ldots,m\}$ into $k$ parts of sizes approximately equal to $m/k$, 
$(\hat{\delta}_{(A_a)}(t), \hat{\eta}_{(A_a)}(t))$ are the estimated parameter functions when the observations in the subset $A_a$ have been left out of the sample, and a local linear maximum likelihood estimator has been used with bandwidth $h$. 

Finally, it is worth mentioning that the local constant maximum likelihood estimation 
\citep{AcarDelicado:2025:IWFOS} has also been implemented in a similar way. 
All the source code used for the present paper can be found in GitHub at
\url{https://github.com/pedrodelicado/LocalLikelihoodBeta}.

\subsection{A toy example}\label{sec:toy_example}
\begin{figure}
   \centering
    \includegraphics[width=1\linewidth]{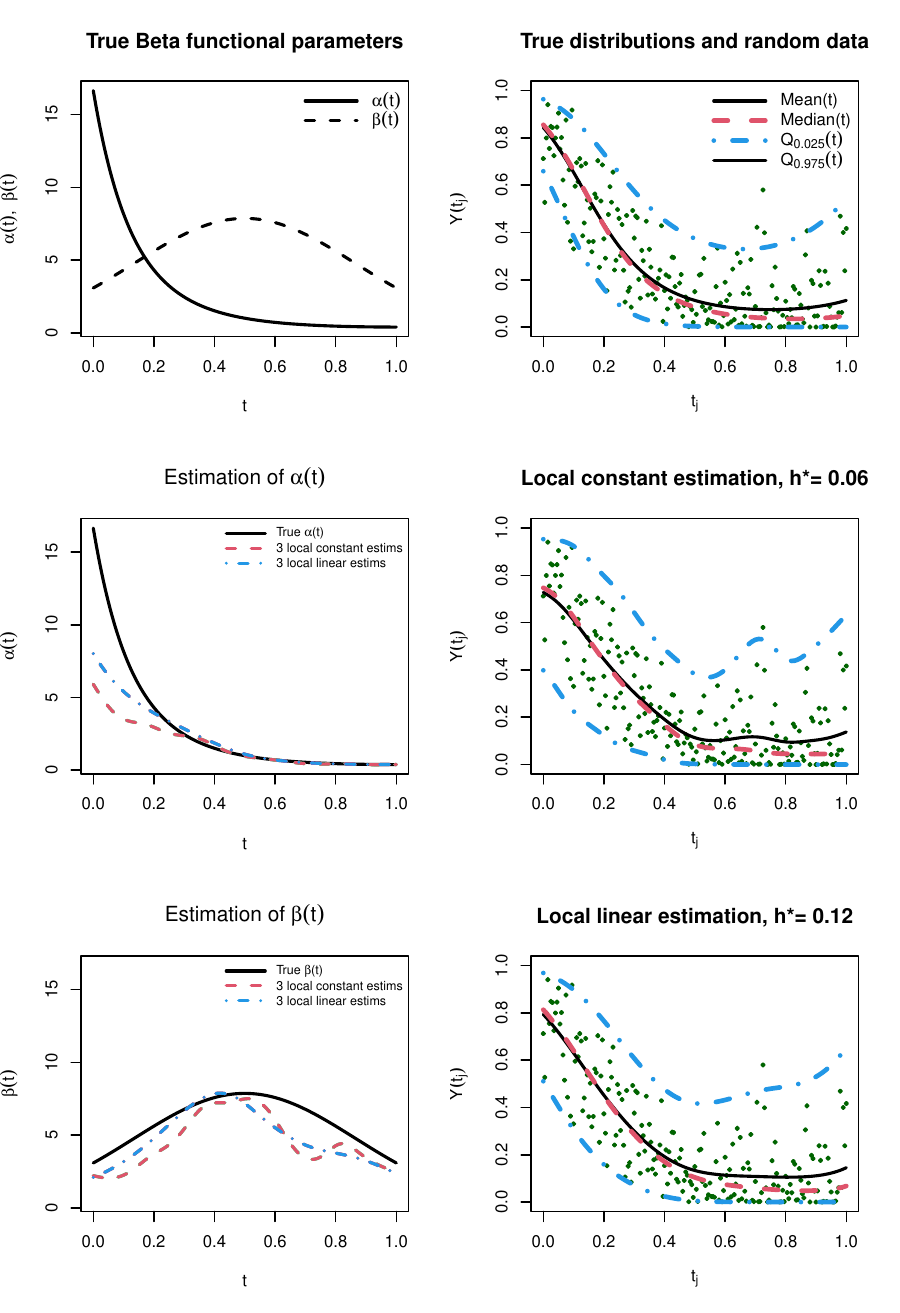}
\vspace*{-1cm}
    \caption{A toy example. A Beta functional model with functional parameters $\alpha(t)$ and $\beta(t)$ (top left panel) is considered. A total of $m=201$ independent observations are generated (top right panel). 
    Local constant and local linear maximum likelihood estimators of the functional parameters $\alpha(t)$ and $\beta(t)$ (middle and bottom left panels).
    The estimated Beta functional models are shown in middle and bottom right panels.   
    }
    \label{fig:toy_example}
\end{figure}

To check the performance of the implemented estimation methods, a toy example has been created. 
For $t\in[0,1]$, we define 
\[
\delta(t)=\frac{15}{4}\left( \left(t-1\right)^2 - \frac{1}{4} \right),
\, 
\eta(t)=-\frac{15}{4}\left( \left( t - \frac{1}{2}\right)^2-\frac{11}{20} \right),
\]
$\alpha(t)=\exp(\delta(t))$, and 
$\beta(t)=\exp(\eta(t))$. 
These last two functions are shown in the top left panel of Figure \ref{fig:toy_example}.
We consider the Beta functional model
$Y(t)\sim \mbox{Beta}(\alpha(t),\beta(t))$, $t\in [0,1]$.
The top right panel of Figure \ref{fig:toy_example} represents the mean, median, and two quantiles of $Y(t)$ as functions of $t$. 

Let $m=201$ and let $Y_j$, $j=0,\ldots, (m-1)$, be independent observations from $\mbox{Beta}(\alpha(t_j),\beta(t_j))$ where $t_j=j/(m-1)$. 
The randomly generated points are shown in the top right panel of Figure \ref{fig:toy_example}.

Six different strategies were considered for estimating functions $\alpha(t)$ and $\beta(t)$:
Local constant or local linear maximum likelihood estimation, each of them using one of three optimization alternatives (the function {\tt optim} with the {\tt Nelder-Mead} or {\tt BFGS} options, and the function {\tt nlm}).


The middle and bottom left panels of Figure \ref{fig:toy_example} show the estimated parameter functions evaluated on a regular grid of 101 points
(bandwidths have been chosen as explained later in this section). 
The three local constant estimators are indistinguishable from each other. The same is true for the three local linear fits, which resemble the true parameter functions more closely than local constant procedures do.

Regarding computing times, we have compared them using the R library {\tt microbenchmark} \citep{microbenchmark}  
and the median running times (in milliseconds) been obtained in 10 repetitions are shown in Table \ref{tab:mb_optimizers}.
It can be seen that local constant maximum likelihood estimation is always faster than local linear, mainly when the function {\tt nlm} is used. 
When using local linear based estimation, the option {\tt BFGS} of function {\tt optim} and the function {\tt nlm} requires very similar times. In a similar experiment with $m=1001$ (not reported here) it was observed that {\tt optim} with {\tt BFGS} was significantly faster than {\tt nlm}.
Taking into account these findings, from now on we will use {\tt nlm} for local constant likelihood estimation, and {\tt optim} with {\tt BFGS} for local linear cases.

\begin{table}
\begin{center}
    \begin{tabular}{l ccc}
    & {\tt optim}
    & {\tt optim}
    &  \\
    & {\tt (Nelder-Mead)}
    & {\tt (BFGS)}
    & {\tt nlm} \\
Local constant max.lik.  &  244 &  88 &  51 \\
Local linear max.lik.    & 2162 & 505 & 492
    \end{tabular}
    \caption{Comparison of median running times of optimization functions {\tt optim} (with methods {\tt Nelder-Mead} and {\tt BFGS}) and {\tt nlm} for estimating local constant and local linear maximum likelihood estimators.}
    \label{tab:mb_optimizers}
\end{center}
\end{table}

A critical aspect of local likelihood estimation methods is the bandwidth choice, which significantly impacts the quality of the estimate and the computational efficiency.
To explore this last aspect, we have used the R library {\tt microbenchmark} to compare the computation times of 
the three implemented methods:
leave-one-out cross-validation (loo), 
approximate leave-one-out cross-validation (approx-loo),
and $k$-fold cross-validation with $k=5$.
Table \ref{tab:mb_bandwidth_choice} shows the median running times (in milliseconds) obtained in 10 repetitions.

\begin{table}
\begin{center}
    \begin{tabular}{l ccc}
    & {\tt loo}
    & {\tt approx-loo}
    & {\tt 5-fold} \\
Local constant max.lik.  &  666 &  2241 &  500 \\
Local linear max.lik.    & 9627 & 11610 & 9643
    \end{tabular}
    \caption{Computation times of the methods leave-one-out cross-validation ({\tt loo}), approximate leave-one-out cross-validation ({\tt approx-loo}), and k-fold cross-validation with k = 5 to select bandwidth ({\tt 5-fold}).}
    \label{tab:mb_bandwidth_choice}
\end{center}
\end{table}

The most notable finding of this exercise is that the approximate leave-one-out cross-validation method requires greater computation times than the na\"ive leave-one-out or the $k$-fold approaches to cross-validation (with a slight advantage for the latter). 
This outcome goes against our initial intuition, which was supported by the proposal of \cite{Loader:1999} for distributions with only one functional parameter, as well as by the statistical tradition of model selection via methods that approximate out-of-sample behavior, such as AIC or BIC.

Although the practical performance was somewhat disappointing, we consider the theoretical developments carried out in Section \ref{sec:local_likelihood_two}, which conclude with Theorem \ref{th:CV}, to be valuable in and of themselves. First, they have allowed us to identify a nonparametric functional estimation problem for which na\"ive cross-validation is more efficient than an elegant, influence-function-based approximation. We firmly believe that publishing negative research results is valuable, as acknowledged, for instance, by \cite{mehta2019Nature} or \cite{Pei_et_al:2024:Clinics}. Second, the analytical developments in Section \ref{sec:local_likelihood_two} have allowed us to obtain expressions for the gradient and Hessian of the maximized function. This considerably facilitates the work of numerical optimization routines. 

Based on the findings in this section, we will use 5-fold cross-validation whenever a bandwidth choice is necessary.
Using this procedure, we find the optimal values to be $h^* = 0.06$ for local constant maximum likelihood estimation and $h^* = 0.12$ for the local linear counterpart in our toy data example. 
The corresponding estimated Beta distributions are shown in the middle and bottom right panels of Figure \ref{fig:toy_example} (local constant on the left and local linear on the right), through the estimated mean, median, and two quantile functions. As can be seen, both estimators resemble the true distributions (top right panel of Figure \ref{fig:toy_example}), with the local linear approach yielding slightly better results.


\section{A real data application} \label{sec:App}
\begin{figure}
\begin{center}
\hspace*{-1.75cm}
\includegraphics[scale=.8]{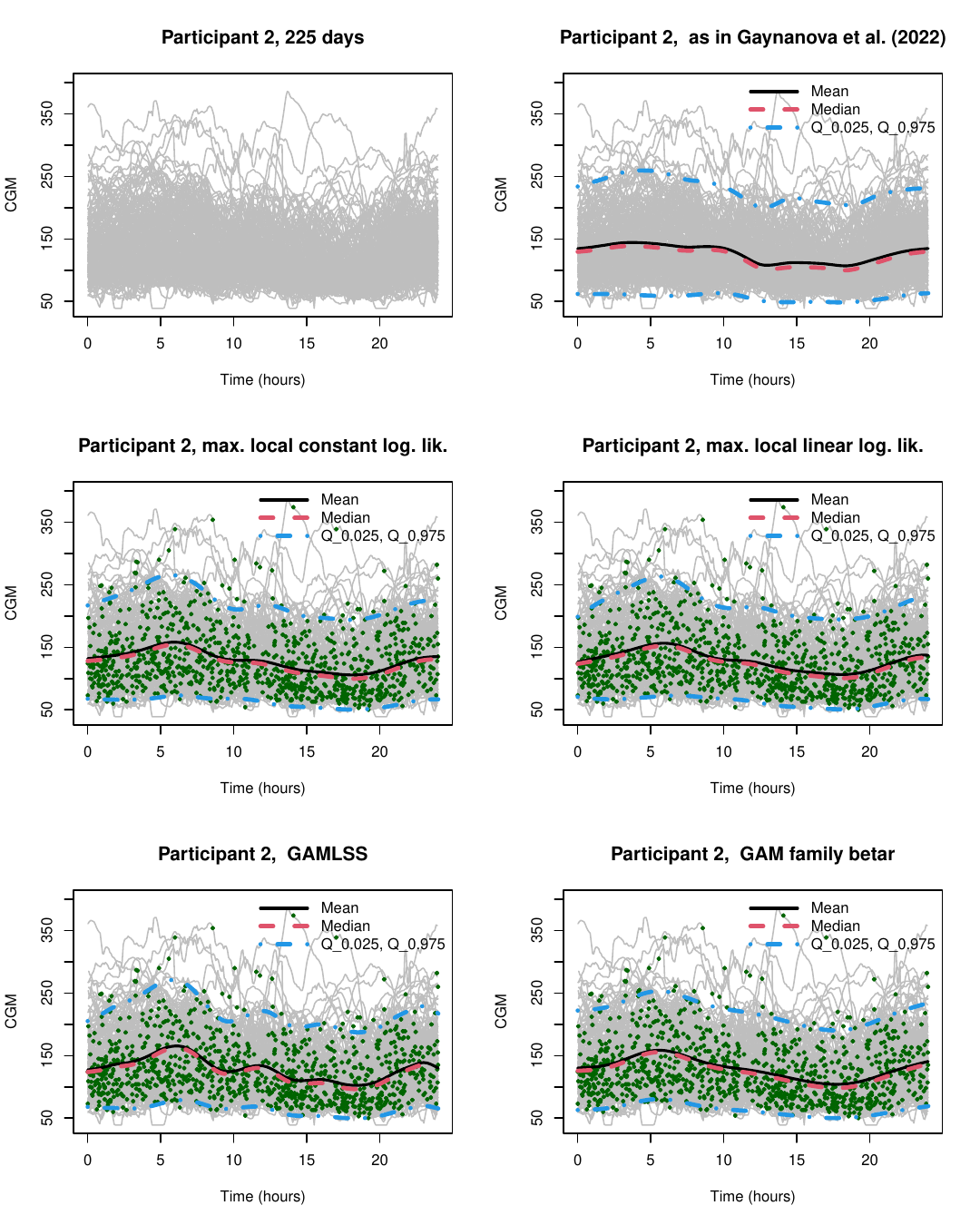}
\end{center}
\caption{CGM curves corresponding to participant ``2'', and estimation of the multilevel Beta functional model by five different methods.}
    \label{fig:CGMs_2_5_estims}
\end{figure}

We applied proposed method to a real data set from the randomized clinical trial study, ``REPLACE-BG" \citep{aleppo2017replace}.
The sample consisted of $n=226$ adult patients (aged over 18 years) who had been diagnosed with type 1 diabetes (T1D) for at least one year at 14 different centers that participated in the T1D Exchange Clinic Network. 
The participants' glucose concentrations were measured by using Dexcom G4 Platinum CGM device. The registration frequency was one observation every 5 minutes (12 observations per hour, 288 per day).
However, some observations were missing. 
We remove days with few observations (less than 220) or with significant gaps (an interval without observations of at least three hours) from the study. 
Then, the average observed number of days per participant is 212.5 days (minimum 48 days, median 214.5 days, maximum 297 days).
For these days, there are on average 270.4 observations (11.3 per hour).
On average, there are 57460 ($212.5\times 270.4$) observations per participant.
As an example, the top left panel of Figure \ref{fig:CGMs_2_5_estims} shows the 225 CGM curves recorded for one of the participants (identified as participant ``2'').


We have fitted the Beta functional model by five different methods:
\begin{enumerate}
    \item \cite{Gaynanova2022}, as explained in Section \ref{sec:MlBeta_GayPunCra}.
    \item Maximum local constant log-likelihood \citep{AcarDelicado:2025:IWFOS}. 
    \item Maximum local linear log-likelihood (see Section \ref{sec:loc_lik_est_Beta}).
    \item GAMLSS, as implemented in function {\tt gamlss} from the library {\tt gamlss} (see Section \ref{sec:MlBeta_simplified}).
    We use family {\tt BEo}, which uses the original parameterization of the Beta distribution ($\alpha(t), \beta(t)$). The two  extra parameters of {\tt gamlss} (skewness $\nu$ and kurtosis $\tau$) have not been considered in the fitting process.
    \item GAM with a Beta distribution, as implemented in function {\tt gam} from the library {\tt mgcv} (see Section \ref{sec:MlBeta_simplified}). Specifically, the family {\tt betar} is used. This choice allows the mean of the Beta distribution $\mu$ to depend on time $t$, while the variance is estimated as $\mu(t)(1-\mu(t))/(1+\phi)$, where the parameter $\phi$ is constant on $t$. 
    Therefore, this approach is less flexible than the previous four, given than only one functional parameter, $\mu(t)$, is estimated. 
\end{enumerate}

For practical reasons, we have unified the observation times of all CGM curves to every 15 minutes from 0 to 24 hours when applying the proposal of \cite{Gaynanova2022}. This has been done by linear interpolation. Using 15-minute intervals instead of the original 5-minute intervals reduces computing load in terms of time and memory.

We have written our own implementation of the estimation methods proposed by \cite{Gaynanova2022}. 
To avoid degeneracies when evaluating Beta densities at $0$ or $1$,
the minimum $m_i$ and the maximum $M_i$ for each individual $i$ are taken, respectively, as $0.99\cdot\min_{k,v} Y_{ik}(s_v)$ and 
$1.01\cdot\max_{k,v} Y_{ik}(s_v)$.
The FPCA steps have been done using the function {\tt fpca.face} from package {\tt refund} \citealp{refund:2024}.
The number of knots used in the spline representations have been fixed to 15, and the proportion of variance explained to 0.9. 
The top right panel of Figure \ref{fig:CGMs_2_5_estims}
shows the results for participant ``2''.
The mean, the median and the quantiles 0.025 and 0.975, are represented as functions of time.


Regarding the other four fitting methods, observe that they assume independent observations in the multilevel Beta functional model (Section \ref{sec:MlBeta_simplified}).
To approximately meet this assumption, $m$ observations are selected at random for each individual, among all days and instant times recorded for that individual. 
This way, independence between any two selected data is guaranteed if they correspond to different days. For pairs of data from the same day, independence is nearly achieved if the observation times are far enough apart. We hope that this is the case when the data are chosen at random. 
We have chosen $m=1000$ (considerably much lower than the average number of observations managed to fit \citealp{Gaynanova2022}, which is 57460).
Thus, on average, there will be fewer than five selected data points corresponding to the same day, with each point about four hours apart. 

The maximum local constant and local linear log-likelihood fits have been implemented with the following specifications.
Estimations are made for times located every 15 minutes between 0 and 24 hours (as in our implementation of \citealp{Gaynanova2022}).
The bandwidth $h$ is chosen by 5-fold cross-validation, maximizing the average out-of-sample log-likelihood, in 10-minute increments from 1 to 2 hours.
The middle panels of Figure \ref{fig:CGMs_2_5_estims}
show the results of the local constant (left) and local linear (right) fits for participant ``2''.



When fitting a GAMLSS with the function {\tt gamlss}, we use smoothing splines to estimate parameter functions, $\alpha(t)$ and $\beta(t)$,  as smooth functions of time. The degree of smoothing is controlled by the {\em degrees of freedom} of the smoothing splines, which we choose by $5$-fold cross-validation (using the function {\tt gamlssCV} from package {\tt gamlss}) from 5 to 25 in increments of 5. 
The estimated parameter functions are evaluated at times located every 15 minutes between 0 and 24 hours (as in the previous estimation approaches). 
The bottom left panel of Figure \ref{fig:CGMs_2_5_estims} shows the results of the GAMLSS fit for participant ``2''.


Finaly, the GAM model estimation is done with the default parameters of the function {\tt gam} with family {\tt betar}, as it is implemented at package {\tt mgcv}. In particular, the choice of tuning parameters controlling smoothness is done by generalized cross-validation.  
The estimated parameter functions are evaluated at times located every 15 minutes between 0 and 24 hours (as before). 
The bottom right panel of Figure \ref{fig:CGMs_2_5_estims} shows the results of the GAM fit for participant ``2''.





In order to compare the quality of the different methodologies, 
we compute the average of the log-likelihood of the five fitted models evaluated 
at all the available data which were not used in the fitting process of the last four models (those using only $m=1000$ observations for each participant in the study).
This way, for each individual in the study, we have an approximation of the out-of-sample performance of these four models.
For the case of the model of \cite{Gaynanova2022} all the points were used in the fitting process, but we compute the same average log-likelihood quantities also for this method. 

\begin{figure}[t]
    \centering
\includegraphics[width=0.85\linewidth]{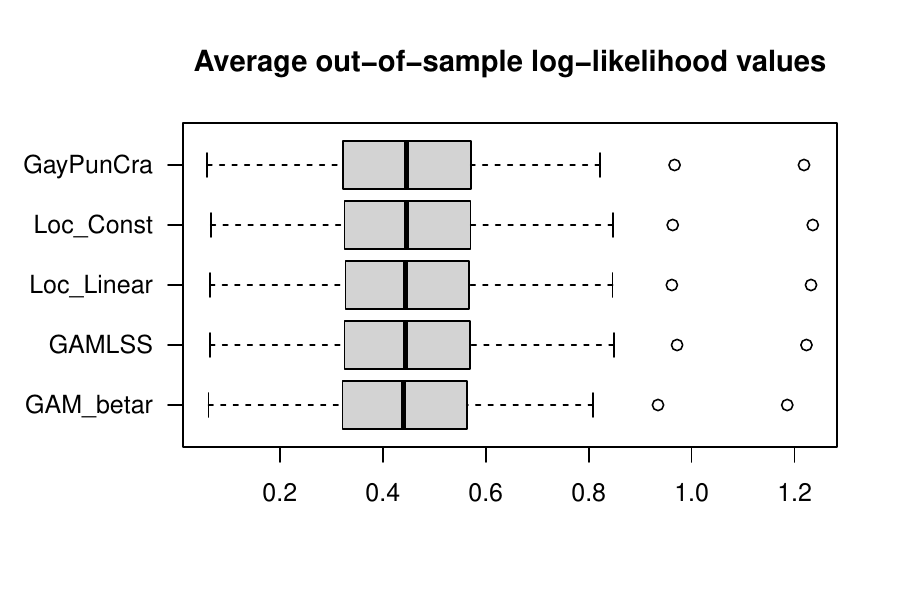}
\vspace*{-1cm}
    \caption{Box-plot of the 226 individual average out-of-sample log-likelihood values for five different estimation methods.}
    \label{fig:boxplots_5_methods}
\end{figure}

Figure \ref{fig:boxplots_5_methods} shows the box-plot of the 226 individual average out-of-sample log-likelihood values for the five different estimation methods, and Table \ref{tab:mean_of_means} lists the means of the 226 individual average out-of-sample log-likelihood values. 
You can see that the five methods give very similar results, with a small advantage in mean for local constant maximum likelihood fit.

\begin{table}[t]
    \centering
    \begin{tabular}{ccccc}
 GayPunCra & LocConst & LocLinear &  GAMLSS & GAMbetar \\ 
   0.4461  &  0.4492  &   0.4478  &  0.4480 &  0.4390 
    \end{tabular}
    \caption{Mean of the 226 individual average out-of-sample log-likelihood values for five different estimation methods.}
    \label{tab:mean_of_means}
\end{table}

In order to determine if the mean differences are significant or not, we have done paired-sample t-tests for the results of each pair of fitting methods. Each test consider two paired vectors of 226 observations. We test equality of means against the two-side alternative hypothesis of different means. 
Table \ref{tab:t_tests} shows the t-tests results. Each entry in this table corresponds to the t-test comparing the corresponding column and row estimation method.
The signs ($-$ or $+$) are those of the difference of the column minus the row methods, while 
the numbers are the $p$-values of the corresponding t-tests.

\begin{table}[t]
\vspace*{.5cm}
    \centering
    \begin{tabular}{l cccc}
      & GayPunCra & LocConst & LocLinear & GAMLSS  \\ 
LocConst     & $-$ \ 0.000  &         &       &       \\
LocLinear    & $-$ \ 0.019  &  $+$ \ 0.000  &        &      \\
GAMLSS       & $-$ \ 0.013  &  $+$ \ 0.000  &  $-$ \ 0.574 &        \\
GAMbetar     & $+$ \ 0.000  &  $+$ \ 0.000  &  $+$ \ 0.000 &  $+$ \ 0.000  \\
    \end{tabular}
    \caption{Results of paired-sample t-tests that compare the means listed in Table \ref{tab:mean_of_means}. 
    Each entry shows the $p$-value comparing the column and row estimation methods. The sign of the column-minus-row differences  are indicated as $-$ or $+$.}
    \label{tab:t_tests}
\end{table}

From Tables \ref{tab:mean_of_means} and \ref{tab:t_tests} it follows the following ordination between the five different estimation methods:
\begin{center}
LocConst $\succ\succ$ GAMLSS $\approx$ LocLinear $\succ$ GayPunCra $\succ\succ$ GAMbetar.
\end{center}
That is, local constant maximum likelihood clearly outperforms GAMLSS and local linear maximum likelihood (which are comparable with each other), they give slightly better results than the proposal of \cite{Gaynanova2022}, and finally the GAM fit provides the poorest results. 

With respect to the computational performance, the median computational time (in seconds) required for each estimation method to provide the estimated functional Beta model for participant "2," including the tuning parameter choice, is shown in Table \ref{tab:mb_times}. This median time was determined through the execution of 10 runs.
As can be seen, the two methods with the poorest statistical performance (\citealp{Gaynanova2022} and GAM) are much faster than the other three. 
This probably happens because the method proposed by \cite{Gaynanova2022} does not require any optimization steps. Additionally, the {\tt mgcv} R library, which we use for GAM, has been efficiently optimized over the years, particularly with regard to smoothness selection criteria.
Finally, local constant maximum likelihood is faster than GAMLSS and  much faster than local linear maximum likelihood. 

\begin{table}[t]
    \centering
    \begin{tabular}{ccccc}
 GayPunCra & LocConst & LocLinear &  GAMLSS & GAMbetar \\ 
     0.602 &       25 &       115 &      36 &    0.172
    \end{tabular}
    \caption{Median computing time (in seconds) required 
    for five different estimation methods to provide the estimated functional Beta model (including tuning parameter choice) corresponding to participant ``2''. }
    \label{tab:mb_times}
\end{table}

\subsection{From individual estimations to a global view of the dataset}
\label{sec:App_MDS_FPCA}
So far we have obtained estimations for the individual functional parameters $\alpha_i(t)$ and $\beta_i(t)$, $i=1,\ldots,n$, by 5 different methods. One of them \citep{Gaynanova2022} starts from a global approach to the whole dataset and, after reducing the dimensionality by FPCA, comes to individual functional parameter estimations. 

The subsequent four methods proceed with estimation on the strictly individual level. 
In this sub-section we propose a convenient way to joint these individual estimated functional parameters to obtain a global picture of the dataset under study. To be specific, we apply multidimensional scaling to reduce the dimensionality of the estimations. 

Let $\{\widehat{\mbox{Beta}}_i(t)\equiv \mbox{Beta}(\hat{\alpha}_i(t),\hat{\beta}_i(t)):t\in[0,24]\}$ be the estimated
functional Beta distribution estimated for individual $i$ in the sample. 
It can be said that it is a functional object that takes values in the Bayes space of continuous probability distributions
(\citealp{Egozcue_et_al:2006}, \citealp{van_den_Boogaart_et_al:2010}, \citealp{van_den_Boogaart_et_al:2014}).
The Bayes space, which is a Hilbert space, extends multivariate Compositional Data Analysis \citep{Pawlowsky_et_al:2015} to density functions, which can be viewed as infinite-dimensional compositional data. 
The Bayes space has been successfully employed to perform functional data analysis when the functional data are density functions. 
See, for instance, \cite{Delicado:2011}, 
\cite{Hron_et_al:2016} and \cite{Maier_et_al:2025}.

The usual distance in the Bayes space of density functions with support $[a,b]$ is the Aitchison distance, originally defined for finite-dimensional compositional data \citep{Aitchison:1982} and then extended to density functions by \cite{Egozcue_et_al:2006}. For two density functions $f$ and $g$ defined on $[a,b]$, their squared Aitchison distance is 
\[
d_A^2(f,g)=\frac{1}{2(b-a)}\int_a^b \int_a^b 
\left( \log \frac{f(x)}{f(y)} - \log \frac{g(x)}{g(y)}\right)^2 dx dy. 
\]
When $f$ and $g$ belong to the same exponential family with $p$-dimensional natural parameters $\theta_f$ and $\theta_g$ respectively, \cite{Delicado:2011}  shows that $d_A(f,g)=C\|\theta_f-\theta_g\|$, where $C$ is a constant and $\|\cdot\|$ is the Euclidean norm in $\mathbb{R}^p$.
Taking into account that the $\mbox{Beta}(\alpha, \beta)$ distribution is an exponential family with natural parameter $(\alpha, \beta)\in \mathbb{R}^2$, we have that 
\[
d_A^2(\mbox{Beta}(\alpha_1, \beta_1),\mbox{Beta}(\alpha_2, \beta_2)) = C^2 \left( (\alpha_1-\alpha_2)^2 + 
(\beta_1-\beta_2)^2 \right).
\]

The functional objects with which we are dealing are
functional Beta distributions:
$\{\mbox{Beta}(\alpha(t),\beta(t)): t \in [0,24]\}$.
A natural way to define a distance between two functional Beta distributions is by integration of Aitchison distances over the interval $[0,24]$, which we call {\em Intgegrated Aitchison} distance:
\[
d_{\mbox{\scriptsize \em IA}}^2(
\{\mbox{Beta}(\alpha_1(t),\beta_1(t)):t\in[0,24]\},
\{\mbox{Beta}(\alpha_2(t),\beta_2(t)):t\in[0,24]\}
)
=
\]
\[
\int_{0}^{24} d_A(\mbox{Beta}(\alpha_1(t),\beta_1(t)),
\mbox{Beta}(\alpha_2(t),\beta_2(t)))^2 dt
=
\]
\[
C^2 \int_{0}^{24} (\alpha_1(t)-\alpha_2(t))^2 + 
(\beta_1(t)-\beta_2(t))^2 dt
=
\]
\[
C^2 \left( d_{L_2}(\alpha_i,\alpha_j)^2 + d_{L_2}(\beta_i,\beta_j)^2 \right),
\]
where $d_{L_2}$ denotes the $L_2$ distance between square-integrable functions defined on $[0,24]$. 

For $i=1,2$, let $\gamma_i(t)$ be the function concatenating $\alpha_i(t)$ and $\beta_i(t))$:
\[
\gamma_i(t)=
\left\{
\begin{array}{lcl}
	\alpha_i(t) & if & t\in [0,24] \\
	\beta_i(t-24) & if & t\in (24,48]
\end{array}
\right. 
\]
Then the Integrated Aitchinson distance is proportional to the sum of the squared $L_2$-distance between the functions $\gamma_i(t)$:
\[
d_{\mbox{\scriptsize \em IA}}^2(
\{\mbox{Beta}(\alpha_1(t),\beta_1(t)):t\in[0,24]\},
\{\mbox{Beta}(\alpha_2(t),\beta_2(t)):t\in[0,24]\}
)
=
\]
\[
C^2 \left( d_{L_2}(\alpha_1,\alpha_2)^2 + d_{L_2}(\beta_1,\beta_2)^2 \right)
=
C^2 d_{L_2}(\gamma_1,\gamma_2)^2.
\]

\begin{figure}
    \centering
\includegraphics[width=1\linewidth]{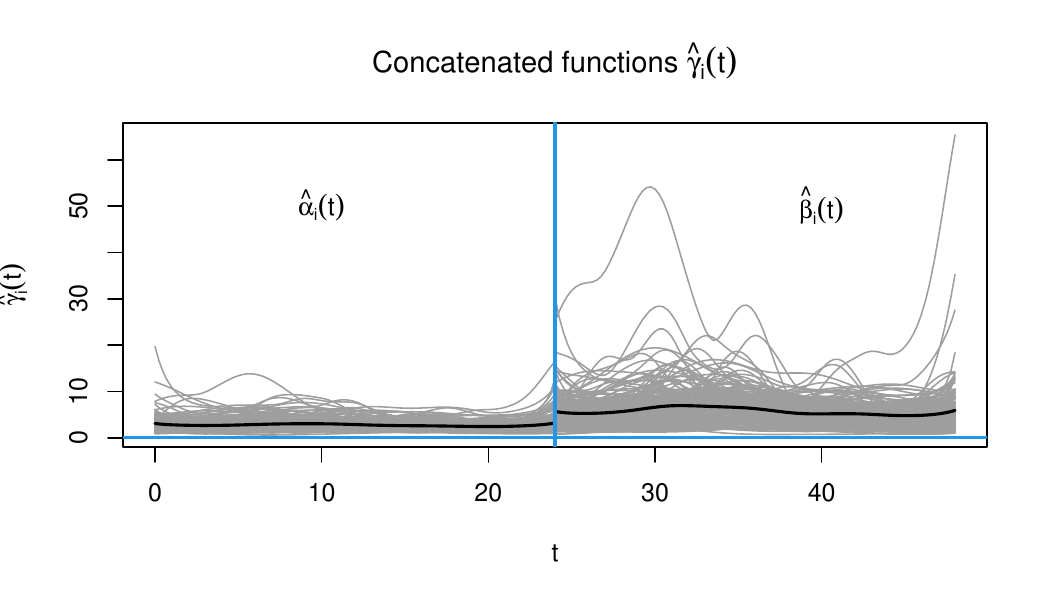}
\\
\vspace*{-.5cm}
\includegraphics[width=1\linewidth]{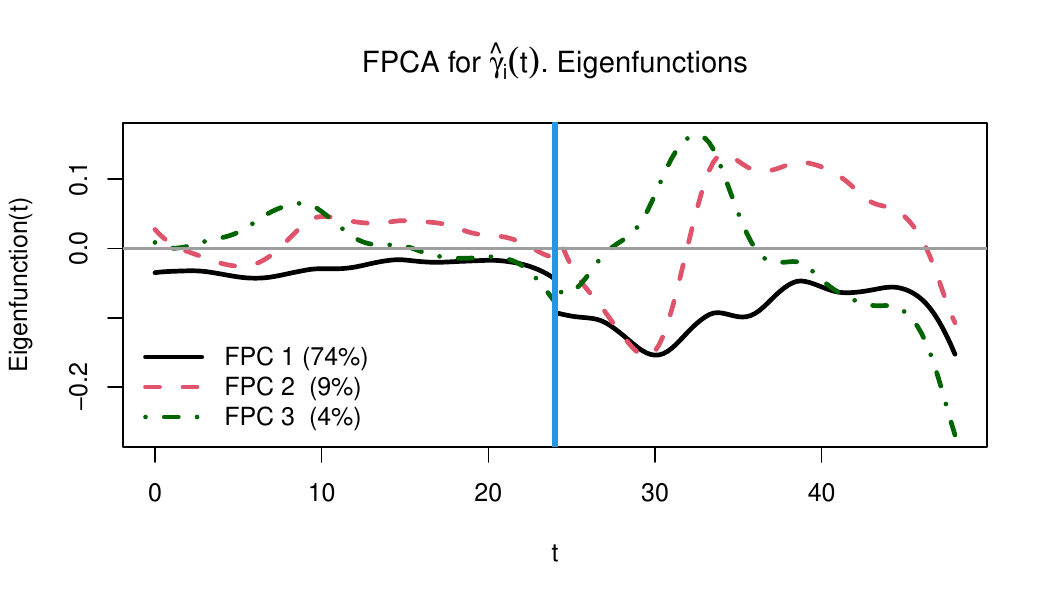}
\vspace*{-1cm}
    \caption{{\em Upper panel:} Concatenated functions $\hat{\gamma}_i(t)$ and their mean function (in black). 
     {\em Lower panel:} FPCA results for functions $\hat{\gamma}_i(t)$.}
    \label{fig:FPCA_gamma_functions}
\end{figure}

We come back to the estimated functional Beta distributions,
\[
\{\widehat{\mbox{Beta}}_i(t)\equiv \mbox{Beta}(\hat{\alpha}_i(t),\hat{\beta}_i(t)):t\in[0,24]\},
i=1,\ldots, n.
\]
We propose to analyze them by MDS based on the Integrated Aitchison distance $d_{\mbox{\scriptsize \em IA}}$. In particular we propose to use classic metric scaling
\citep[Section 14.2]{mardia1979multivariate}. 
Given that the Integrated Aitchison distance coincides (up to a constant) with the $L_2$ distance between the concatenated functions $\gamma$, it follows 
\citep[Section 14.3]{mardia1979multivariate}
that classic metric scaling from $d_{\mbox{\scriptsize \em IA}}$ is equivalent to FPCA applied to the concatenated functions $\hat{\gamma}_i(t)$, $i=1,\ldots, n$,
each being the concatenation of $\hat{\alpha}_i(t)$ and $\hat{\beta}_i(t)$.
The upper panel of Figure \ref{fig:FPCA_gamma_functions} shows the functions $\hat{\gamma}_i(t)$. 

The results of FPCA applied to the functions $\hat{\gamma}_i(t)$ are shown in the lower panel of Figure \ref{fig:FPCA_gamma_functions}. The first 3 principal functions explain 87\% of the total variability. 
The main part of the variability is due to the first functional principal component, which accounts for changes at the global level because the corresponding eigenfunction has a constant sign over all values of $t$. 
The second eigenfunction allows for changes in CGM during sleep time, while the third is responsible for the differences between morning and evening.  

\begin{figure}
    \centering
    \includegraphics[width=1\linewidth]{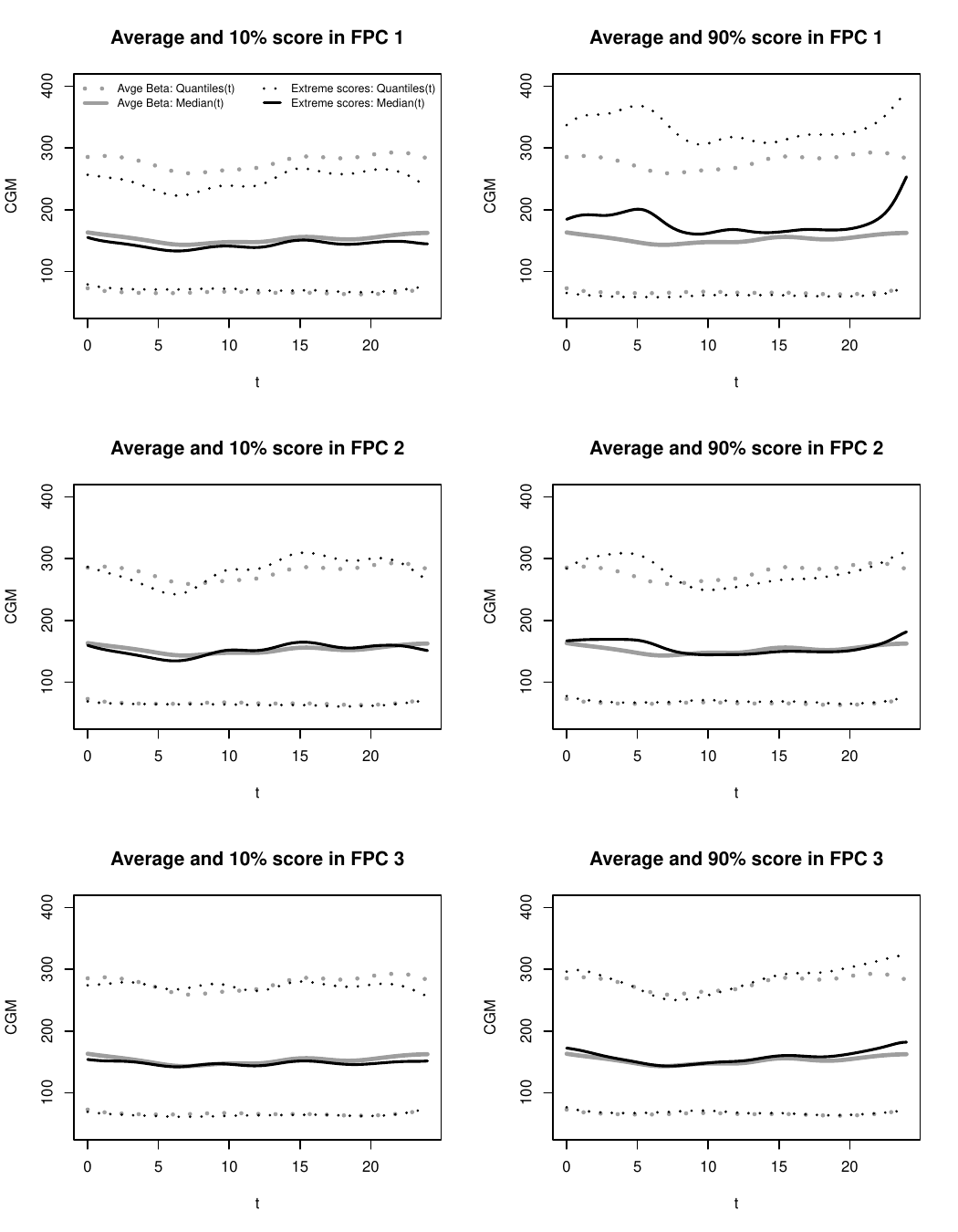}
    \caption{Functional Beta distributions (represented by the median and two 
quantile functions) illustrating the changes in the directions of the first three functional principal components (each one in a row). 
Extreme scores are 10\% or 90\% quantiles (left and right columns).
The average functional Beta distribution is shown in grey in all the panels.
}
    \label{fig:FPCA_output}
\end{figure}

The results of FPCA for functions $\hat{\gamma}_i(t)$, summarized in the lower panel of Figure \ref{fig:FPCA_gamma_functions}, 
are difficult to understand because the scale of the eigenfunctions is that of $\hat{\gamma}_i(t)$ and consequently that of $\hat{\alpha}_i(t)$ and $\hat{\beta}_i(t)$.
To improve interpretability, Figure \ref{fig:FPCA_output} shows the FPCA results in the scale of the observed CGM functions. 
Each panel of this figure compares two functional Beta distributions rescaled to the observed CGM values: 
(1) the average one (in grey), with concatenated functional parameters $\alpha(t)$ and $\beta(t)$ represented in black in the upper panel of Figure \ref{fig:FPCA_gamma_functions};
(2) the functional Beta distribution (in black) corresponding to an extreme quantile (10\% or 90\%, left and right columns of Figure \ref{fig:FPCA_output}, respectively) of the scores in one of the first three functional principal components (each represented in a different row of Figure \ref{fig:FPCA_output}).
Each functional Beta distribution is represented by the median function and quantile functions corresponding to probabilities $0.025$ and $0.975$.
The following observations are derived from the graphics.
The first principal direction differentiates between lower-and-more-concentrated CGM levels over all the day (low scores) and higher-and-more-disperse CGM levels (high scores).
The second principal direction is a contrast between lower and higher CGM levels at sleep time (low and high scores, respectively).
Finally, the third principal direction accounts for changes in CGM levels in the evening and first night hours.

\section{Conclusions}
\label{sec:concl}
This paper has adopted a functional data analysis approach to analyze wearable device data focusing on  CGM functions. 
Based on the model introduced by \cite{Gaynanova2022}, we have proposed a simplification under the assumption that all the observations are independent. 
The proposed model allows individual level estimation using techniques developed for generalized non-parametric models. 
In the literature, this kind of models are generally estimated by using spline based methods such as GAM and GAMLSS where the latter model allows more than one parameter depending on the response. 
Our objective in this study was to explore an alternative approach to GAM and GAMLSS which uses local likelihood estimation instead of splines. In accordance with this purpose, we have extended the local linear maximum likelihood estimation from one to two functional parameters. Moreover, we have developed a theoretical approximation for the bandwidth choice based on leave-one-out cross-validation, following the steps done by \cite{Loader:1999} for the one-parameter case. We have tested the performance of the proposed methods using a synthetic data set. We have found that the local linear estimation slightly outperforms local constant likelihood estimation. However, its computational cost is considerably higher. Additionally, when we compared the approximate leave-one-out cross validation to the na\"ive implementation of leave-one-out and 5-fold cross-validation methods in terms of computation time, the approximate method was not reducing the computational time. 

We have explored a real data set coming from a large clinical trial involving CGM data obtained from wearable devices. The performance of five different estimation methods have been compared. Regarding statistical performance, the five methods gave comparable results, with a small advantage of local constant maximum likelihood, followed by the local linear estimator and the GAMLSS, 
then the method of \cite{Gaynanova2022} and, finally, the GAM method. Regarding computation time, the most efficient were \cite{Gaynanova2022} and GAM methods, followed by local constant maximum likelihood and GAMLSS and, finally, local linear estimation. As a result of all our findings, we recommend to use local constant maximum likelihood with bandwidth choice based on the $k$-fold cross-validation.

In the final step of our analysis of the REPLACE-BG data, we have considered the set of estimated varying-parameter Beta distributions as an abstract  functional data set. 
A dimensionality reduction analysis has allowed us to give a global overview of these data,
extracting common patterns from the individual estimations.

\section*{Acknowledgements}
This research was supported by projects
PID2023-148158OB-I00 (founded by MICIU/AEI/10.13039/501100011033 
and European Union Next Generation 
\linebreak EU/PRTR),
and by UPC under AGRUPS-2024. 
Nihan Acar-Denizli is a Serra-H\'unter fellow.

\section*{Declaration of generative AI and AI-assisted technologies in the manuscript preparation process}

During the preparation of this work, the authors used DeepL in order to check the grammar. After using this tool, the authors reviewed and edited the content as needed and take full responsibility for the content of the published article.

\bibliographystyle{chicago}
\bibliography{Beta_loc_lik}


\appendix
\numberwithin{equation}{section}
\renewcommand{\theequation}{\thesection.\arabic{equation}}
\section{Local likelihood in one-parameter models} 
\label{app:loader}
We summarize here Chapter 4 in \cite{Loader:1999}, devoted to local polynomial likelihood estimation for distribution families with one parameter depending on a explanatory variable. 
For simplicity, we only show results for the local linear fit.

Let $Y_1,\ldots, Y_m$ be independent random variables with probability density (or mass) function $f(y;\theta(t_j))$, $j=1,\ldots, m$, where $0<t_1<\cdots<t_m < 1$ are known constants, 
and $\theta(t)$ is assumed to be a {\em smooth} function from $[0,1]$ to $\mathbb{R}$. 
Let $y_1,\ldots, y_m$ be the observed values of $Y_1,\ldots, Y_m$ at $m$ individuals, for which it is assumed that a certain explanatory variable 
has taken the values $t_1,\ldots, t_m$.

The objective of local likelihood estimation is to estimate the function $\theta(t)$, $t\in[0,1]$, in a non-parametric way.  
Let $t_0$ be the point in $[0,1]$ at which we want to estimate the value $\theta(t_0)$. 
For $t$ in a neighborhood of $t_0$, consider the first-order Taylor approximations

\[
\theta(t)\approx
\theta_{t_0}(t)=\theta(t_0) + \theta'(t_0)(t-t_0) \equiv a_0 + a_1 (t-t_0)
=\langle \bfa, A(t-t_0) \rangle, 
\]
where $\bfa=(a_0, a_1)\tr$ and $A(u)=(1,u)\tr$, for $u\in \mathbb{R}$.
The local linear log-likelihood function around $t_0$ is defined as
\begin{equation}\label{eq:llll_function_1_param}
	\cL_{t_0}(\bfa)=\sum_{j=1}^m w_j(t_0) 
	\ell(y_j;\langle \bfa, A(t_j-t_0) \rangle),
\end{equation}
where $\ell(y;\theta)=\log f(y;\theta)$, 
the weights $w_j(t_0)$ are given by a kernel function $K$ and a bandwidth parameter $h$: $w_j(t_0)=K((t_j-t_0)/h)$. 
Let $\hat{\bfa} =\arg\max_{\bfa} \cL_{t_0}(\bfa)$ be the maximum local linear likelihood estimators of $\bfa$.
Observe that $\hat{\bfa}$ depends on $t_0$. 
The local linear likelihood estimator of $\theta(t_0)$ is defined as
\[
\hat{\theta}(t_0)= 
\hat{\theta}_{t_0}(t_0)=\langle \hat{\bfa}, A(0) \rangle=\hat{a}_0.
\]

The first-order conditions (also known as the estimating equations) for the maximum local linear likelihood problem around a generic $t$ are 
$\nabla_{\bfa} \cL_{t}(\bfa)=\mathbf{0}_2$,
where $\mathbf{0}_d$ is the vector of zeros in $\mathbb{R}^d$. 
Taking into account the definition of $\cL_t$ in equation (\ref{eq:llll_function_1_param}),
\[
\nabla_{\bfa} \cL_{t}(\bfa) = 
\sum_{j=1}^m w_j(t) A(t_j-t) 
\ldottheta(y_j;\langle \bfa, A(t_j-t) \rangle)
=
\bfX\tr \bfW \bfldottheta(\bfX \bfa), 
\] 
where 
$\bfX$ is the $m\times 2$ matrix with $j$-th row $A(t_j-t)\tr=(1,t_j-t)$, 
$\bfW$ is the $m\times m$ diagonal matrix with $j$-th element in the diagonal $w_j(t)$, and 
$\bfldottheta(\bfX \bfa)$ is the column vector in $\mathbb{R}^m$  with $j$-th element 
$\ldottheta(y_j;\langle \bfa, A(t_j-t) \rangle)$,
the partial derivative of $\ell$ with respect to the parameter, evaluated at the parameter value $\langle \bfa, A(t_j-t) \rangle$. 
Therefore, the estimating equations can be written in matrix notation as
$\bfX\tr \bfW \bfldottheta(\bfX \bfa)= \mathbf{0}_2$.

Bandwidth choice is a fundamental issue in local likelihood estimation, as it is the case in all non-parametric estimation methods.
\cite{Aerts_Claeskens:1997} found the asymptotic order of the bandwidth $h$ either minimizing the point-wise asymptotic mean squared error or maximizing the expected log-likelihood of a new observation. These results have no clear translation into practical automatic bandwidth selectors.
They finally propose to choose $h$ as the maximizer of the leave-one-out version of the log-likelihood function, even if they mention the potential large variance of the method. 

For a particular value of the bandwidth $h$, the leave-one-out cross-validation version of the log-likelihood function is 
\begin{equation}\label{eq:loo_log_lik_1_param}
	\text{CV}(h) = \sum_{j=1}^{m} \ell(y_j;\hat{\theta}_{(j)}(t_j)),
\end{equation}
where $\hat{\theta}_{(j)}(t)$ is the estimated parameter function when the $j$-th observation has been left out of the sample and a local linear maximum likelihood estimator has been used with bandwidth $h$. 
Then, $h$ is chosen to minimize $\text{CV}(h)$:
\[
h_{\text{\scriptsize CV}} = \arg \max_h \text{CV}(h).
\]
This bandwidth selector requires to estimate $m$ parameter functions (where $m$ is the number of observed pairs $(t_j, Y_j)$), each of them involving $(m-1)$ maximum local likelihood estimations. 
Therefore, leave-one-out cross-validation is time consuming for large sample sizes.

In certain cases (linear estimators of the regression functions and, in particular, linear smoothers) it is possible to deduce an exact expression for the $j$-th term in (\ref{eq:loo_log_lik_1_param}) from the corresponding one in the full log-likelihood function  
\[
\sum_{j=1}^{m} \ell(y_j;\hat{\theta}(t_j)),
\]
where $\hat{\theta}(t)$ is estimated with the complete data set. 
In this case, the parameter functions have to be estimated only once. 

When exact expression are not available, approximated expressions can help to fastener the computation of (\ref{eq:loo_log_lik_1_param}), as the one presented in \cite{Loader:1999}, which is as follows. 
For $t=t_j$ and for $\lambda\in [0,1]$, consider the modified estimating equations 
\[
\bfX\tr \bfW \bfldottheta(\bfX \bfa)
- \lambda K(0) A(0)  \ldottheta(y_j;\langle \bfa, A(0) \rangle) 
= 
\mathbf{0}_2
\]
and let $\hat{\bfa}(\lambda)$ be the solution. 
It follows that $\hat{\bfa}(0)$ is the maximum local likelihood estimate of 
$\bfa$ at $t=t_j$, and that 
$\hat{\bfa}(1)$ is the leave-one-out counterpart. 
We consider the first-order Taylor approximation of $\hat{\bfa}(1)$ around
$\hat{\bfa}(0)$:
\[
\hat{\bfa}(1)
\approx
\hat{\bfa}(0) + 
\left. 
\frac{d \, \hat{\bfa}(\lambda)}{d\, \lambda}
\right|_{\lambda=0}.
\]
To compute the $2$-dimensional directional derivative with respect to $\lambda$ we write the modified estimating equations evaluated at the solution $\hat{\bfa}(\lambda)$,
\[
\bfX\tr \bfW \bfldottheta(\bfX \hat{\bfa}(\lambda)) 	
=
\lambda K(0)
A(0)  
\ldottheta(y_j;\langle \hat{\bfa}(\lambda), A(0) \rangle),
\]
and take derivatives with respect to $\lambda$ at both sides:
\[
\bfX\tr \bfW 
\bflddottheta(\bfX \hat{\bfa}(\lambda)) \bfX 
\frac{d \, \hat{\bfa}(\lambda)}{d\, \lambda}
=
\]
\[
K(0)
A(0)  
\ldottheta(y_j;\langle \hat{\bfa}(\lambda), A(0) \rangle)
+
\lambda K(0)
{\scriptsize
	A(0) 
	\lddottheta(y_j;\langle \hat{\bfa}(\lambda), A(0) \rangle)  
	\frac{d \, \hat{\bfa}(\lambda)\tr}{d\, \lambda}
	A(0) 
	,
}
\]
where the term $\bflddottheta$ is the $m\times m$ diagonal matrix with second derivatives computed from the elements in $\bfldottheta$ in the diagonal.
Now we evaluate the previous equation at $\lambda=0$ and solve in the 2-dimensional directional derivative. 
To simplify the notation, let us define the $m\times m$ matrix
\[
\bfV=
- 	\bflddottheta(\bfX \hat{\bfa}(0)) 
\]
and the $2\times 2$ matrix 
\[
\bfJ= 
\bfX\tr \bfW \bfV \bfX.
\]
Under the standard assumption that $\bfW\bfX$ has full rank, the concavity of the likelihood function implies that $\bfJ$ is positive definite 
\cite[see][Theorem 4.1]{Loader:1999}, so it is invertible. 
Define also $\bfe_i$, $i=1,2$, as the vector of $\mathbb{R}^2$ with a one
in the $i$-th position and zeros elsewhere.
Then, the derivatives with respect to $\lambda$ of the modified estimating equations, evaluated at $\lambda=0$, can be expressed as
\[
- \bfJ 
\left. 
\frac{d \, \hat{\bfa}(0)}{d\, \lambda}
\right|_{\lambda=0}
=
K(0) 
\bfe_1
\ldottheta(y_j;\langle \hat{\bfa}(0), A(0) \rangle),
\]
and it follows that
\[
\left. 
\frac{d \, \hat{\bfa}(0)}{d\, \lambda}
\right|_{\lambda=0}
=
- K(0) 
\bfJ^{-1} 
\bfe_1
\ldottheta(y_j;\langle \hat{\bfa}(0), A(0) \rangle).
\]
Therefore,
\[
\hat{\theta}_{(j)}(t_j) 
=
\bfe_1\tr \hat{\bfa}(1) 
\approx 
\bfe_1\tr \hat{\bfa}(0)
-
K(0) 
\bfe_1\tr \bfJ^{-1} \bfe_1
\ldottheta(y_j;\langle \hat{\bfa}(0), A(0) \rangle)
=
\]
\[
\hat{\theta}(t_j) 
-
K(0) 
\bfe_1\tr \bfJ^{-1} \bfe_1
\ldottheta(y_j;\hat{\theta}(t_j)).
\]
\cite{Loader:1999} defines the influence function at $t_j$ as 
$\infl(t_j) = K(0) \bfe_1\tr \bfJ^{-1} \bfe_1$.
Then,
\[
\hat{\theta}_{(j)}(t_j)
\approx 
\hat{\theta}(t_j)
-
\infl(t_j)
\ldottheta(y_j;\hat{\theta}(t_j)).
\] 
Now we approximate $\ell(y_j;\hat{\theta}_{(j)}(t_j))$ with a first-order Taylor expansion:
\[
\ell(y_j;\hat{\theta}_{(j)}(t_j))
\approx 
\ell(y_j;\hat{\theta}(t_j))
-
\infl(t_j)
\ldottheta(y_j;\hat{\theta}(t_j),\hat{\eta}(t_j))^2.
\] 
Therefore,	
\[
\text{CV}(h) = 
\sum_{j=1}^{m}
\ell(y_j;\hat{\delta}_{(j)}(t_j))
\approx 
\]
\[
\widetilde{\text{CV}}(h) = 
\sum_{j=1}^{m} 
\ell(y_j;\hat{\delta}(t_j))
- 
\sum_{j=1}^{m}
\infl(t_j)
\ldottheta(y_j;\hat{\theta}(t_j),\hat{\eta}(t_j))^2.
\]
As a consequence, the bandwidth $h$ can be chosen to maximize $\widetilde{\text{CV}}(h)$:
\[
h_{\widetilde{\text{\scriptsize CV}}} = \arg \max_h \widetilde{\text{CV}}(h).
\]
Taking into account that the deviance is defined as minus twice the 
log-likelihood of the fitted model, the deviance computed from the leave-one-out fitted model is
\[
\mathrm{Dev}_{\mathrm loo}(h)
\approx  
\mathrm{Dev}(h) 
+  
2 \sum_{j=1}^{m}
\infl(t_j)
\ldottheta(y_j;\hat{\theta}(t_j))^2.
\]
\cite{Loader:1999} defines the {\em effective degrees of freedom} of the estimation as
\[
\nu =
\sum_{j=1}^{m}
\infl(t_j)
\ldottheta(y_j;\hat{\theta}(t_j))^2,
\]
Then, a generalization of the Akaike information criterion follows:
\[
\mathrm{AIC}(h)=\mathrm{Dev}(h) 
+ 2 \nu.
\]
Therefore, 
\[
h_{\widetilde{\text{\scriptsize CV}}} = 
\arg \max_h \widetilde{\text{CV}}(h) =
\arg \min_h \mathrm{AIC}(h).
\]

\end{document}